\documentclass[11pt, a4paper]{article}
\pdfoutput=1

\usepackage{jheppub}
\usepackage{mathtools}
\usepackage{latexsym}
\usepackage{appendix}
\usepackage{mathrsfs}
\usepackage{graphicx}
\usepackage{microtype}
\usepackage{feynmp-auto}
\usepackage[export]{adjustbox}
\usepackage{enumitem}
\usepackage{relsize}
\usepackage{afterpage}

\usepackage{simpler-wick}

\usepackage{shuffle}

\def\be{\begin{equation}}
\def\ee{\end{equation}}
\def\nn{\nonumber}

\def\R{\mathbb{R}}

\def\I{\mathcal{I}}
\def\J{\mathcal{J}}

\renewcommand{\Re}{\operatorname{Re}}
\renewcommand{\Im}{\operatorname{Im}}
\renewcommand{\ln}{\log}

\usepackage{xspace}

\def\bkrules{Bern:1990cu,Bern:1990ux,Bern:1991aq,Bern:1993wt}

\title{Loop amplitudes monodromy relations and color-kinematics duality}

\author[a]{Eduardo Casali,}\emailAdd{ecasali@ucdavis.edu}
\affiliation[{a}]{Center for Quantum Mathematics and Physics (QMAP) and\\
Department of Physics, University of California, Davis, CA 95616 USA}

\author[{b}]{Sebastian Mizera,}\emailAdd{smizera@ias.edu}
\affiliation[{b}]{Institute for Advanced Study, Einstein Drive, Princeton, NJ 08540, USA}

\author[{c}]{Piotr Tourkine}\emailAdd{piotr.tourkine@lpthe.jussieu.fr}
\affiliation[{c}]{Laboratoire de Physique des Hautes \'Energies, CNRS \& Sorbonne Universit\'e, 4 Place Jussieu, 75005 Paris, France}

\abstract{Color-kinematics duality is a remarkable conjectured property of gauge theory which, together with double copy, is at the heart of a wealth of new developments in scattering amplitudes.
So far, its validity has been verified in most cases only empirically, with limited ab initio understanding beyond tree-level.
In this paper we provide initial steps in a first-principle understanding of color-kinematics duality and double-copy at loop level, through a detailed analysis of the field-theory 
limit of the monodromy relations of string theory at one loop.
In this limit, we dissect the type of Feynman graphs  generated and the relations they obey.
We find that graphs with contact-terms are unavoidable and are generated in the field theory limit of ``bulk'' contours which do not have a standard physical interpretation in string perturbation theory. We show how they are related to ambiguities in the definition of the loop momentum and that their role is precisely to cancel those ambiguities.
}

\begin{document}
\maketitle

\section{Introduction}
\label{sec:introduction}

The color-kinematics duality is a conjectured property of the perturbative expansion of gauge theory amplitudes 
proposed by Bern, Carrasco, and Johansson (BCJ) \cite{Bern:2008qj}.
It was born as a means of constructing gravity amplitudes via the {double-copy} procedure \cite{Bern:2010ue}. The range of application of these techniques have been remarkably wide, from amplitudes to classical solutions of General Relativity and gravitational wave emission patterns, to string theory. A comprehensive review can be found in \cite{Bern:2019prr}.

It is therefore all the more remarkable that the property at the root of these developments, the color-kinematics duality, is  known to hold to arbitrary multiplicity for tree-level processes \cite{BjerrumBohr:2009rd,Stieberger:2009hq,Feng:2010my,Bern:2010yg}. In particular, it is still a conjectured property of loop amplitudes, and it is even less clear how it is implemented at the level of non-linear classical solutions. If the conjecture can be proven true at higher loops, it would not only be very useful in simplifying computations of scattering amplitudes, but would also reflect a deep relationship between perturbative gauge theories and quantum gravity, invisible at the level of their respective Lagrangians.

This duality has been extensively checked for amplitudes at loop orders with a bottleneck at five loops~\cite{Bern:2017yxu,Bern:2017ucb,Bern:2018jmv}. Despite its many successes, we remain completely ignorant as to whether or not the duality continues to hold true or as to how it should be applied in a completely general setting.

Our approach to this problem, which has proven useful in the past, will be to use string theory. At tree-level, the color-kinematics duality is indeed fully understood from string theory. It originates from fundamental identities in open-string theory scattering amplitudes, known since the early days of dual models~\cite{Plahte:1970wy}, today known as \emph{monodromy relations}~\cite{Plahte:1970wy,BjerrumBohr:2010hn,BjerrumBohr:2010zs,BjerrumBohr:2009rd,Stieberger:2009hq}.
Those relations were generalized to loop-level in \cite{Tourkine:2016bak,Hohenegger:2017kqy}, and recently seen to emanate from a deeper mathematical structure known as  {twisted homology}~\cite{Mizera:2017cqs,Mizera:2017rqa,Mizera:2019gea,Casali:2019ihm,Mizera:2019blq}. 

Over the past few years a related approach based on ambitwistor string theory has emerged, see, e.g., \cite{Geyer:2015bja,Geyer:2015jch,Geyer:2016wjx,Geyer:2017ela,Geyer:2019hnn,Edison:2020uzf,Farrow:2020voh} (various other recent worldsheet approaches to color-kinematics duality include \cite{Mafra:2011kj,Ochirov:2013xba,He:2015wgf,Mafra:2017ioj,Fu:2018hpu,Fu:2020frx,Mafra:2018pll,Gerken:2020yii,Gerken:2019cxz,Edison:2020ehu}), which gives a handle on the problem of constructing BCJ numerators, at a cost of introducing linearized propagators which need to be transformed into quadratic ones using non-trivial partial fraction identities. Despite many successes of this research direction, our goal here is to obtain Feynman diagrams directly from worldsheet degenerations, which at present is understood most appropriately in the case of string theory.

\textbf{Mysterious transverse integrals in the monodromy relations:}
These relations, however, revealed another conundrum:
In open string theory, gauge bosons are represented by strings with color charges at their ends, the Chan-Paton factors. This implies that vertex operators of gluons are always inserted at the \textit{boundaries} of open-string worldsheets.
The mysterious feature of the monodromy relations and their associated twisted cycles at loop-level is that they involve integrating the vertex operators of gluons \textit{into the bulk} of the worldsheet. From the perspective of string theory, this is an exotic phenomenon, which, presently, has no physical interpretation.
In \cite{Tourkine:2016bak,Tourkine:2019ukp,Casali:2019ihm} it was suggested that they are related to the color-kinematics duality, but this statement was not made precise.

\textbf{The labeling problem in the color-kinematics duality:}
A Yang-Mills amplitude can be written as an expansion involving only trivalent Feynman diagrams by expanding the four-point vertex for instance. In $d$ space-time dimensions, the $n$-gluon Yang-Mills amplitude at the $L$-th loop order is then written as
\begin{equation}
  \int \prod_{i=1}^L\frac{d^d\ell_i}{(2\pi)^{ d}}
  \underbrace
  {\sum_{\substack{\mathrm{trivalent}\\ \mathrm{graphs}\, \Gamma}} \frac{1}{S_\Gamma} \frac{n_\Gamma\, c_\Gamma}{D_\Gamma}}_{\textstyle =\I(\ell_1,\dots,\ell_L) }.
  \label{eq:Int-YM}
\end{equation}
Contributions from each trivalent graph features kinematic numerators $n_\Gamma$, which depend on external and internal kinematics; the color factors $c_\Gamma$, which are products of structure constants $f^{abc}$; and products of Feynman propagators $D_\Gamma$ associated to this specific graph. A symmetry factor $1/S_\Gamma$ also needs to be inserted.

Color-kinematics duality states that, given all triples of color factors $c_\Gamma$'s obeying Jacobi identities of the form
\begin{equation}
\includegraphics[valign=c]{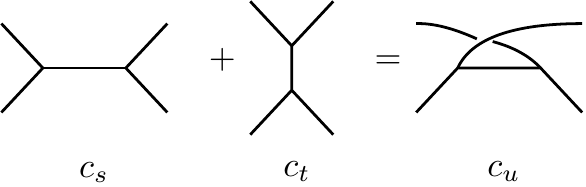}
\end{equation}
there exists a representation of the amplitude where the kinematic numerators $n_\Gamma$ also satisfy the same Jacobi identities.
When this representation exists, fewer kinematic numerators have to be computed, e.g. planar graphs relate to non-planar graphs: this reduces the complexity of computing the amplitude.

However, this representation suffers from some ambiguities. A natural one is the possibility to shift the numerators by quantities that vanish in a Jacobi identity. This is a structural ambiguity akin to gauge redundancy. A more severe ambiguity, and one we address in the text in our framework, comes from the freedom of redefining loop momenta in field theory. This means that a notion of ``the'' integrand  $\I(\ell_1,\dots,\ell_L)$ as in eq.~\label{sec:introduction-1}~\eqref{eq:Int-YM} is usually ill-defined.

\begin{figure}
  \centering
  \includegraphics{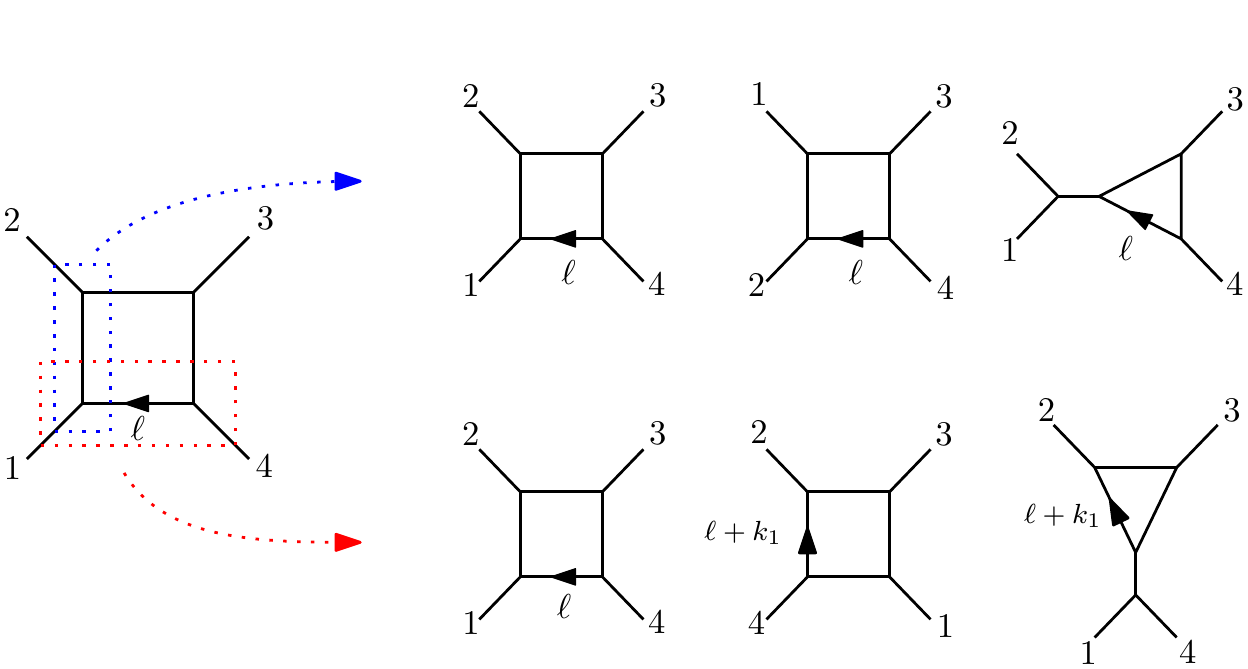}
  \caption{Illustration of the labeling problem. At one-loop in string theory the loop momentum can be globally defined by the property that it always starts after the $n$-th leg, here leg $4$ on the left-hand side box graph.
    \underline{Top:} example of BCJ identity which does not change the definition of the loop momentum. \underline{Bottom:} identity which changes the definition of the loop momentum. Note that the rightmost graph has a correctly defined loop momentum because leg $1$ is to the left of leg $4$.}
  \label{fig:BCJ-labeling}
\end{figure}

In contrast, string theory has a well defined notion of the integrand, on which a global definition of loop momentum can be introduced using the formalism of chiral splitting~\cite{DHoker:1988pdl,Tourkine:2019ukp}. It is then likely that following this notion of integrand the through the field-theory (or tropical \cite{Tourkine:2013rda}) limit gives, if not a canonical, at least a ``nice'' representation for a field theory integrand. Figure~\ref{fig:BCJ-labeling} illustrates this problem in the case of $n=4$ particles. The labeling induced by string theory is that the loop momentum always starts after leg $4$: this is a gauge choice coming from fixing translation invariance on the annulus. The problem is that there are Jacobi identities which exchange the position of this leg and modify the definition of the loop momentum in mismatching ways.

In field theory, one is able to cook up a solution and declare that the numerator of the mismatched graph is equal to that of the other, but at higher loop order this question become more tricky. This phenomenon is called the \textit{labeling problem} and is actually one of the bottleneck in finding color-kinematics satisfying representations. There are no rules to determine which graphs should be used at higher loop orders, e.g. no rule to tell if graphs with different labeling of internal loop momenta should count as different graphs with different numerators or not.

One goal of this paper is to use in the field-theory limit of string theory monodromy relations to see what string theory has to say about this question.

\paragraph{Summary of the results:}
\label{sec:plan-paper}

\begin{itemize}[leftmargin=*]
\item We find that the field theory limit of the monodromy relations produces numerators which automatically satisfy Jacobi identities inside the graph, i.e., in those places where the definition of the loop momentum would not be changed by a Jacobi move, as explained above.
\item We characterize the extra contributions arising from the bulk transverse integrals of the annulus present in the monodromy relations. We carefully compute their field theory limit and show that it produces two types of graphs: contact terms, and graphs with trees attached to the loop. The existence of the first class was suggested in \cite{Ochirov:2017jby}, but the second are completely new. These graphs enter the monodromy relations in a crucial way by removing the graphs where a Jacobi identity would be ambiguous otherwise, in the sense that it would require a cancellation between two graphs with different definitions of the loop momentum. Therefore, string theory evades the problem of loop-momentum redefinition by effectively \textit{removing the ambiguous identities}.
\end{itemize}

We would like to add that it is not our intention to imply that monodromy relations lead to BCJ-satisfying numerators. In particular, the stringy way to solve the monodromies, as we detail in this text, does not produce BCJ identities at those points where the loop momentum jumps, and rather adds contact terms so as to satisfy the monodromy relations.

In the discussion section we elaborate on the significance of these results in the context of gravity. Contact terms to be squared seem in particular unavoidable, which furnishes an a posteriori justification for the generalized double-copy procedure of \cite{Bern:2017yxu,Bern:2017ucb}. This also hints towards the physical role of the bulk integrals as a possible new underlying structure in the color-kinematics duality.

The paper is organized as follows. In section~\ref{sec:reviews}, we review the mechanism of the field theory limit and the monodromy relations. In section~\ref{sec:field-theory-limit}, we describe the field theory limit of the bulk contours and how they generate contact terms and triangle-type graphs. This can be seen as a new item in the Bern-Kosower rules, required for the monodromy relations. In section~\ref{sec:field-theory-limit-monodromy} we show how the field theory limit of the monodromy relations produces numerators which satisfy BCJ identities in the bulk, and how the bulk contours remove the terms in which the BCJ identities could have been spoiled by redefinitions of the loop momentum. We summarize the paper in section~\ref{sec:discussion}, where we also comment on the extensions of to higher-loop orders and interpretation of bulk cycles in the context of double-copy.

\section{Reviews of the tropical limit and monodromy relations}
\label{sec:reviews}

\subsection{Field-theory limit and Bern-Kosower rules}
\label{subsec:ftl}
In this section, we present a short review of the field-theory limit of open-string theory \cite{Bern:1990cu,Bern:1990ux,Bern:1991aq,Bern:1993wt}. Field theory amplitudes are generated by sending $\alpha'\to0$ in a string amplitude, more precisely $\alpha' k_i\cdot k_j\ll 1$ for all $i,j$. We take all external states to be massless, $k_i^2 =0$. In the absence of UV divergences, the leading order contributions to this amplitude, after suitable rescaling, become Feynman graphs.\footnote{When there are UV divergences, it is sufficient, for our purposes, to truncate the modulus integrations in the amplitudes, as our relations are valid pointwise in the moduli space. This results in Schwinger proper-time amplitudes with a hard cut-off of order $\alpha'$ for the Schwinger proper-time.}
This scaling limit can be also understood as coming from a tropicalization of the moduli space of punctured Riemann surfaces~\cite{Tourkine:2013rda}, therefore in the text we will use the terms \emph{tropical limit} and \emph{field-theory limit} interchangeably. We refer to~\cite{Tourkine:2013rda} for conventions, signs and factors of $\pi$ and $2$'s which are necessary for a clean analysis of the limit. It is crucial to keep track of these factors given how delicate some cancellations are.

In the field theory limit, the moduli space integration of string theory only receives contributions from regions near its boundaries, corresponding to the Riemann surface degenerating into graphs with different topologies. Intuitively, the open-string worldsheet becomes a collection of infinitely long and thin ribbons, with widths proportional to $\sqrt{\alpha'}$, joining and splitting at interaction points. The resulting object depends only on the length of the edges which correspond to Schwinger proper-time parametrization of Feynman graphs after suitable rescaling. At one-loop, this process is systematized by the Bern-Kosower rules~\cite{\bkrules}. We refer the reader to \cite{Schubert:2001he} for a thorough review, and recall below only the aspects of these rules necessary for our purposes.
For concreteness, we will focus on the one-loop case but the basic idea generalizes to all genera: we comment on the higher-loop case in the discussion section~\ref{sec:discussion}.

\begin{figure}
	\begin{center}
		\includegraphics[scale=0.8]{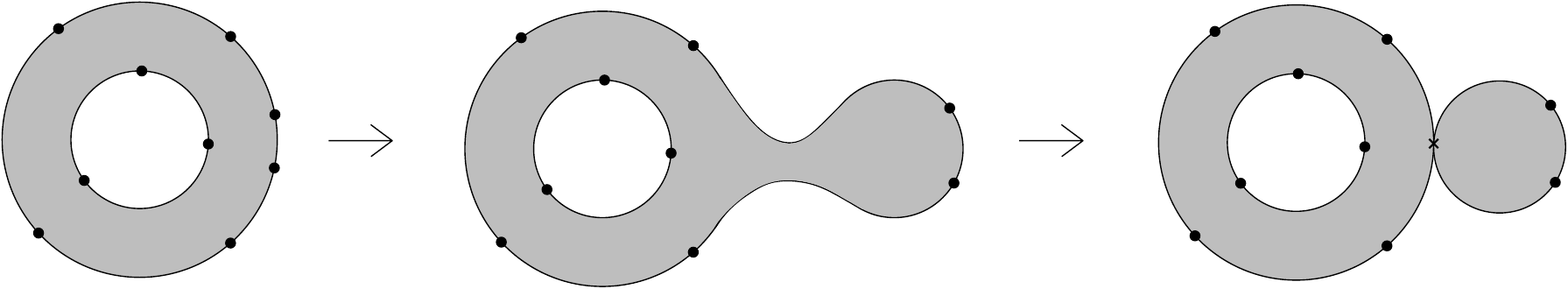}
	\end{center}
	\caption{Example of a separating degeneration at one-loop.}
	\label{fig:sep_deg}
\end{figure}

\begin{figure}
	\begin{center}
		\includegraphics[scale=0.8]{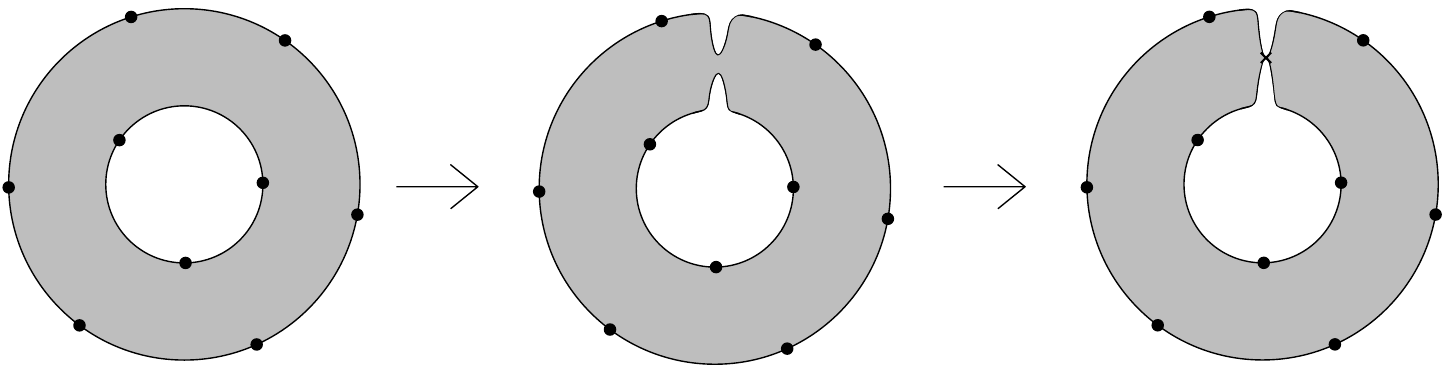}
	\end{center}
	\caption{Example of a non-separating degeneration at one-loop.}
	\label{fig:non_sep_deg}
      \end{figure}

There are two types of degenerations at the boundaries of the moduli space: \textit{separating} and \textit{non-separating}. A separating degeneration occurs when the original surface pinches and splits into two surfaces connected at a point, or equivalently by an infinitely long strip, see figure~\ref{fig:sep_deg}.
A non-separating degeneration occurs when the pinched surface is a connected surface with a double-point, see figure~\ref{fig:non_sep_deg}.
As an example, take a one-loop open-string amplitude with $n$ ordered punctures on the same worldsheet boundary. Its field theory limit generates all possible trivalent graphs that have this ordering: the $n$-gon, and all other one-loop graphs with trees attached to the loop. The attached trees are generated from boundary components where two or more punctures get very close together and the worldsheet pinches as depicted on the right of figure \ref{fig:sep_deg}.
The generic case of a $g$-loop graph with particles ordered on the $g+1$ boundaries obey the same mechanism. Therefore, all graphs which respect a given ordering are generated in the field-theory limit.

However, this does not mean that a given string amplitude has support on all of these graphs. For instance, supersymmetry can prevent the appearance of certain graphs, such as triangles in maximally supersymmetric theories, see~\cite{Bern:1998sv,Bern:2005bb,BjerrumBohr:2005xx,BjerrumBohr:2006yw,Bern:2007xj,BjerrumBohr:2008ji}. What happens in this case is that the string integrand has zero support on those degenerations at
leading order in $\alpha'$.

What properties of a string integrand indicate whether or not it has support on a given boundary of the moduli space? To answer this question, we specialize to one-loop, but the statements below are generic since they depend only on the local structure of the propagator and not the topology (genus) of the surface. A typical string integrand assumes the following form
\begin{equation}
  \label{eq:typical-string}
  \varphi(\{z_i\})\, \times e^{-\alpha' \ell^2 \Im\tau -2\pi \alpha' \sum_{i=1}^n \ell\cdot k_i \Im z_i +\sum_{i<j}k_i \cdot k_j G_{ij}},
\end{equation}
where the exponent, which is traditionally called Koba-Nielsen factor, is universal to all string amplitudes, and $\varphi$ is a theory-dependent function with no branch cuts. The annulus is defined by a rectangle of height $t$ and width $1/2$, so that $\tau=it$. The punctures $z_i$ live on both boundaries, $0\leq \Im z_i\leq t$ and $\Re z_i=0,1/2$. We follow the conventions of \cite{Tourkine:2016bak,Casali:2019ihm}. The Koba-Nielsen factor is constructed out of the following function, \footnote{It differs from the Green's function  $\langle X(z_i) X(z_j) \rangle$ by a non-holomorphic term proportional to $(\Im z_{ij})^2/\Im \tau$. We always compensate this term by working in the chiral splitting formalism\cite{DHoker:1988pdl}, which introduces a loop momentum integration. Consequently, we always work at fixed loop momentum, i.e. before integration.
}
\begin{equation}
  G_{ij}:=G(z_i -z_j)=-\alpha'\ln \bigg| \frac{\vartheta_1(z_i-z_j)}{\vartheta_1'(0)} \bigg| \,
  \label{eq:corr-XX}
\end{equation}
explicitly given by
\begin{equation}
  \label{eq:G-def}
  G(z)/\alpha'=
  -\ln | \sin(\pi z) | +  
  2\sum_{m=1}^\infty \frac{q^m}{1-q^m}\frac{1}{m}\cos(2\pi m z) +c(\tau)
  \,.
\end{equation}
Here $\vartheta_1$ is the first Jacobi theta function, $q = e^{2\pi i\tau}$ is the nome of the Riemann surface with modular parameter $\tau$ and $c(\tau)$ is a function that eventually drops out of computations due to momentum conservation. For the annulus we have $\tau \in i\R$; for more conventions see appendix~\ref{sec:conventions}.
The monodromy properties of the integrand are entirely given by this factor, which contains all the branch cuts of the integrand. In other words, it controls the monodromy structure regardless of the matter content in a specific scattering process.

The remaining part of the integrand, $\varphi$, is naively given by a polynomial in the derivatives of $G$ and kinematic constant terms (powers of internal and external momenta). Bern and Kosower proved that it is always possible to find a sequence of integrations by part so that the $\varphi$ has only first derivatives of $G$~\cite{\bkrules}, see also the review \cite{Schubert:2001he}.\footnote{This reasoning is valid at one-loop. Possible obstructions at higher loop involve the risk that integration by parts may interact non-trivially with picture changing operators.} 
Therefore, the function $\varphi$ can always be written as a polynomial in $\dot{G}_{ij}$, and takes the most general form:
\begin{equation}
  \label{eq:schem}
  \varphi(\{z_i\}) = \sum_\alpha c_\alpha \prod_{(i,j)\in \alpha} \dot G_{ij},
\end{equation}
where $\alpha$ is a set of pairs of labels $(i,j)$ appearing in a given term and $c_\alpha$'s contain the polarization and kinematic dependence of the amplitudes. Note that $n$ powers of $\dot G$ correspond to a numerator with $n$ powers of the loop momentum in the field theory limit.

Now, following Bern and Kosower consider the pair $(i,j) \in \alpha$. We ask whether a given monomial $\varphi_\alpha = \prod_{(i,j)\in \alpha} \dot G_{ij}$ splits off a tree or not. There are two cases: 1) $\dot G_{ij}$ appears with two or no powers, i.e., $\dot G_{ij}^{2} \subset \varphi_\alpha$ or $\dot G_{ij} \not\subset \varphi_\alpha$, or 2) exactly one power of $\dot G_{ij}$ appears in $\varphi_\alpha$. The mechanism of the field theory limit, systematized by the Bern-Kosower rules, stipulates that case 1 gives an integrand with no support on graphs where legs $(i,j)$ forms an external tree, while case 2 gives has support on those graphs, as well on other graphs, where $(i,j)$ do not split off a tree.

\paragraph{Case 1: no (ij)-tree.}

In the field theory limit, the annulus becomes infinitely long, so that $\tau\to i\infty$, $z_j\to i \infty$ with a tropical scaling
\begin{equation}
  \Im z_j = \frac{Y_j}{\pi \alpha'}\,,\qquad\Im \tau=\frac T {\pi\alpha'}\,,
  \label{eq:trop-scal}
\end{equation}
The quantities $Y_j$ and $T$ are the field theory Schwinger proper-times of the graph. 
The propagator reduces to
\begin{equation}
  \label{eq:trop-G}
  G_{ij}= -\ln(|\sinh(Y_j-Y_i)/\alpha' |) = -|Y_j-Y_i|/\alpha' +
  {\cal O}(e^{-2 |Y_{ij}|/\alpha'}),
\end{equation}
where terms with non-zero powers of $q$ are exponentially suppressed in the field theory limit and drop out.\footnote{The story is more complicated than this and depends on the amount of supersymmetry. The string partition function may possess terms of order $q^{-1}$ or $q^{-1/2}$ which extract residues at order  $q^{1}$ or  $q^{1/2}$. The effect of these terms, fully systematized in the original Bern-Kosower rules, is to adapt the number of powers of $\dot G_{ij}$ in the numerator to the amount of SUSY and the spin of the particles. It does not change the fact that the integrand is solely made of powers of $\dot G_{ij}$.}
Equation \eqref{eq:trop-G} approaches, as expected, the worldline propagator 
$-|Y_j -Y_i|$
in the limit $\alpha'\rightarrow 0$, when taking into account the factor of $\alpha'$ in eq.~\eqref{eq:typical-string}.

\paragraph{Case 2: (ij)-tree.}

A tree graph occurs when a separating degeneration pinches off a punctured disk from the original surface, or equivalently when a set of punctures comes infinitesimally close to each other. Consider the case where two particles, $z_i$ and $z_j$, approach each other such that a three-punctured disk splits off. The answer to our question above is that a string integrand will have support on this degeneration if $\varphi_\alpha$ contains exactly one power of $\dot G_{ij}$.

In the region $z_i-z_j\ll1$, the integrand can then be written by as 
\begin{equation}
\varphi(\{z_i\})\, e^{\sum_{r,s}k_r \cdot k_s G_{rs}}=\dot G_{ij}e^{k_i \cdot k_j G_{ij}}\times \left(\tilde \varphi\, e^{\sum_{r,s\neq i} k_r \cdot k_s G_{rs}}\right)\bigg|_{z_i=z_j}\,+{\cal O}(z_i-z_j),
\end{equation}
where the Bern-Kosower rules stipulate that the ${\cal O}(z_i-z_j)$ terms drop out in the field theory limit.
Thus, the only part of the integrand which still depends on the variable $z_i$ is $\dot G_{ij}e^{k_i \cdot k_j G_{ij}}$.
From the derivative of the Green's function,
\begin{equation}
  \label{eq:G-dot-def}
  \partial_z G(z) =-\frac{\vartheta_1(z)'}{\vartheta_1(z)} =
  -\pi \cot(\pi z)- 4\pi
  \sum_{m=1}^\infty \frac{q^m}{1-q^m}\sin(2\pi m z)\,
\end{equation}
and from $G_{ij}$, we retain only the first term in the $q$-expansion, as the ${\cal O}(q)$ terms are exponentially suppressed in the limit \eqref{eq:trop-scal}.
Then, on an integration contour where $y_i=\Im z_i$ approaches $y_j=\Im z_j$ from below, we perform the tropical limit rescaling and zoom around the piece of the contour near $y_j$. For a term involving $\dot G_{ij}e^{k_i \cdot k_j G_{ij}}$ this gives us 
\begin{align} 
  \I_{\text{trop}} &= i \int_{y_j-L}^{y_j}dy_i \cot(i\pi(y_j-y_i)) e^{\alpha' k_1\cdot
    k_j\ln(-i\sin(i\pi(y_j-y_i)))} \nn \\
&= \frac{(\sinh (\pi L )) ^{\alpha'
      k_1\cdot k_j}}{\pi \alpha' k_1\cdot k_j}  = \frac{1}{\pi \alpha' 
    k_1\cdot k_j}
  +{\cal O}(1),\label{eq:tr-usual}
\end{align}
where $L$ is some cut-off which drops out in the limit. Besides, since $L$ was inserted by hand, the full integral cannot depend on it order-by-order in $\alpha'$, it must therefore vanish when taking into account the other parts of the contour, where $y_i$ is above $y_j$. An explicit example is given in eqs~\eqref{eq:tr-usual-1}, \eqref{eq:tr-usual-2}, where we see that the dependence on $L$ is pushed to $O(\alpha')^2$ compared to leading order.

On the right hand side of \eqref{eq:tr-usual} we recognize immediately the propagator of a tree subgraph with legs $i$ and $j$. If particle $y_i$ approaches from above the result is the same apart from an overall minus sign from the antisymmetry of the cotangent function. After integrating out $y_i$ in this way the rest of the integrand is given by the previous integrand with $\dot G_{ji}$ removed and $z_i$ replaced everywhere by $z_j$. In section~\ref{sec:field-theory-limit}, we will redo this calculation to extract the field theory limit of the bulk contours entering the monodromy relations. Because the standard string amplitudes do not involve those contours, their analysis was absent from the older literature on the field theory limit of strings.

\subsection{Monodromy relations}

Monodromy relations \cite{BjerrumBohr:2009rd,BjerrumBohr:2010zs,Stieberger:2009hq} are linear relations between open-string theory amplitudes known to exist at tree-level since the early days of dual models \cite{Plahte:1970wy}. These relations can be used to solve for a basis of the BCJ color-kinematics duality, see references above and also \cite{Feng:2010my}. They were extended to loop level in \cite{Tourkine:2016bak,Hohenegger:2017kqy,Ochirov:2017jby}, and formalized in the context of twisted homologies by the present authors in \cite{Casali:2019ihm}. The reader should refer to~\cite{Casali:2019ihm} for more details, conventions and proofs of the identities used in this paper.

Mathematically, monodromy relations are linear relations between integrals over the configurations space of points on sliced genus-$g$ Riemann surfaces, whose integrand involves a multi-valued function, the Koba-Nielsen factor, $T_g$.
At tree-level, this function is given by
\begin{align}
 T_0(\{z_1,\dots,z_n\})=\prod_{i<j}(z_j-z_i)^{\alpha' k_i\cdot k_j},
\end{align}
where $k_i$ is the momentum associated with puncture $z_i$.
The monodromy relations at tree-level can be expressed as relations among color-ordered open-string \textit{amplitudes}, where a single puncture circulates around, starting from its original position. Taking the ordering $12\dots n$ and circulating $1$ for instance gives Plahte's relations~\cite{Plahte:1970wy}:
\begin{align}
  \sum_{i=1}^{n-1}e^{\pm \pi i\alpha' k_1\cdot \sum_{j=2}^ik_j}A_{\text{tree}}(2,\dots, i, 1, i{+}1,\dots,n ) =0,
  \label{eq:plahte}
\end{align}
where $A_\text{tree}(1,\dots,n)$ denotes tree-level open string amplitudes in a particular color ordering. These are two separate relations labeled by a sign $\pm$.

\begin{figure}\begin{center}
 \includegraphics{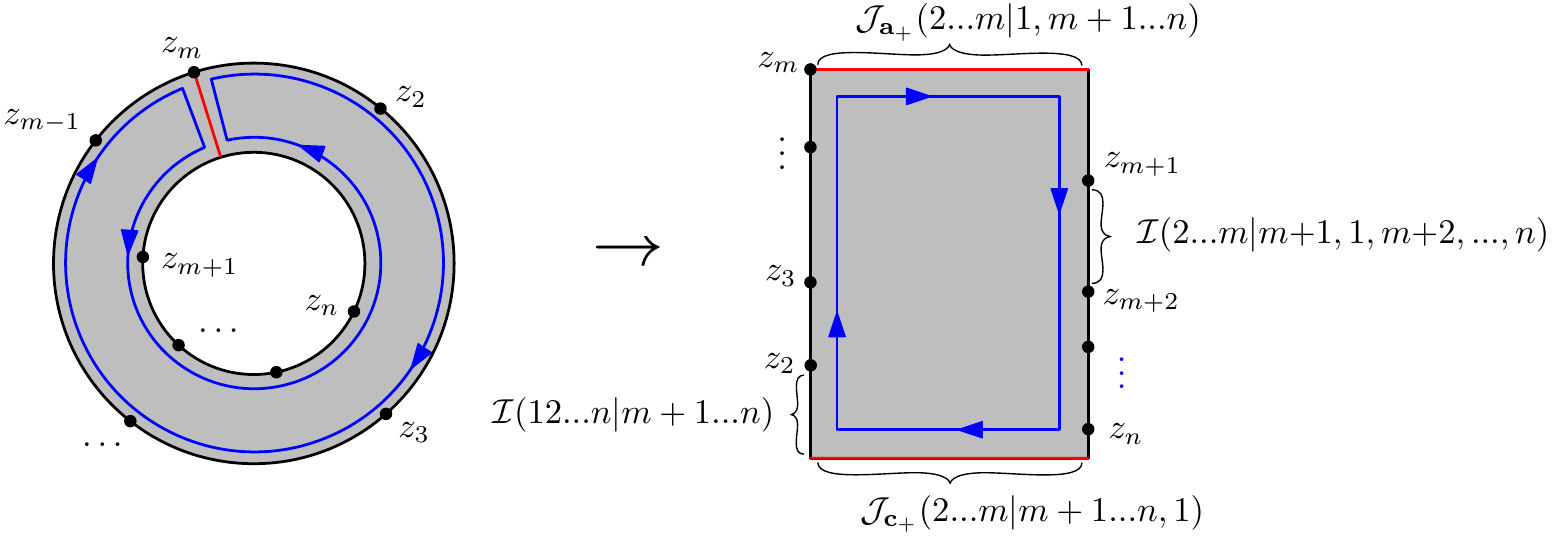}
 \end{center}
 \caption{Illustration of monodromies relations coming from the vanishing of an integral around the closed blue contour,cv for a generic non-planar amplitude. \underline{Left:} open string annulus, with punctures. Red line: $A$-cycle, along which the Riemann surface is cut, defines where the loop momentum $\ell^\mu = \int_A \partial X^\mu$ is measured. Blue cycle: the contour over which the puncture $z_1$ is being integrated. As no pole exist in the bulk, the full integral vanishes. Each segment along the boundary. \underline{Right:} rectangle representation of the annulus, with depictions of contours for $z_1$. The $\mathcal{I}$ contours are usual open string boundary contours, the $\mathcal{J}$ are bulk contours.}
 \label{contours_def}
\end{figure}

At loop-level, the monodromy relations can be expressed as relations between color-ordered open-string \textit{loop integrands}: they hold at fixed surface moduli and fixed loop momenta. Most of the contributions to these relations are integrated over the usual open-string cycles, i.e., the particles are ordered along the $g{+}1$ boundaries of a Riemann surface, in accordance with a given Chan-Paton ordering.
But there are also unavoidable contributions from \emph{bulk cycles}, coming from contours that run in the interior of the surface, along its $A$-cycles, see, e.g., the red lines in figure~\ref{contours_def}.
It is worth recalling that, so far, they have no interpretation as originating from the string theory path-integral.

At genus one, fixing $m{-}1$ punctures on one boundary and $n{-}m{-}1$ on the other (we fix $z_m=it$ by translation invariance), the monodromy relations can be written as
\begin{align}\label{string_non_planar}
&\qquad\sum_{i=1}^{m-1}e^{\pm\pi i \alpha' k_1\cdot \sum_{j=2}^{i} k_{j}} \,\I(2,3,\dots, i,1,i{+}1,\dots, m | m{+}1,\dots, n)\nn\\
&\qquad\qquad\qquad +\sum_{i=m}^n e^{\pm\pi i \alpha' k_1\cdot(\ell+ \sum_{j=2}^{i}k_{j})} \,\I(2,\dots, m|m{+}1, \dots, i, 1, i{+}1,\dots, n)\\
&\!= \mp e^{\pm\pi i \alpha' k_1 \cdot \ell} \!\left( e^{\pm\pi i \alpha' k_1 \cdot \sum_{j=2}^{m} k_j}\J_{\mathbf{a}_\pm}(2,\dots, m|1,m{+}1,\dots, n) {-} \J_{\mathbf{c}_\pm}(2,\dots, m| m{+}1, \dots, n,1) \right)\!,\nn
\end{align}
where $\I (\cdots|\cdots)$ denote a physical integration contour with the two slots denoting the ordering of punctures on each boundary, and $\mathcal{J_{\mathbf{a}_{\pm}}}$, $\mathcal{J_{\mathbf{c}_{\pm}}}$ denotes the contributions integrated along $A$-cycles as denoted in figure~\ref{contours_def}.
Those are the relations we use in this paper. The relation with minus signs in the phases and $\mathcal{J}_{\mathbf{a_-/c_-}}$ is obtained by drawing the same vanishing contour, but on a reflected rectangle, see \cite[Figure 9]{Casali:2019ihm}.

The general form of the $\mathcal{J}$ terms is
\begin{multline}
 \J_{\mathbf{a/c}_{\pm}}(2,\dots,m|1,m{+}1,\dots,n)=\int_{\Delta} \prod_{i\neq 1,m}d z_i e^{-2\pi \alpha' \ell\cdot \sum_{i\neq 1}^nk_i \Im(z_i)}\\\times\prod_{i,j\neq 1}|G(z_i,z_j)|^{\alpha' k_i\cdot k_j}\int_0^{\pm1/2} d x_1 T_1(z_1),
\end{multline}
where $x_1=\Re z_1$. The integration contours are $\Im z_1=0$ for $\mathcal{J}_{\mathbf{c}_\pm}$ and $\Im z_1=t$ for $\mathcal{J}_{\mathbf{a}_\pm}$. The contour $\Delta$ is the usual one for the $n{-}2$ punctures distributed along the two boundaries and we have fixed the $m$-th puncture to $z_m=\tau=it$. The function $T_1(z_1)$ is obtained from analytic continuation of the string integrand in the variable $z_1$,
\begin{equation}\label{T1}
T_1(z_1) := e^{2\pi i \alpha' k_1\cdot \ell\, z_1} \prod_{j=2}^m \left(-i \vartheta_1(iy_j{-}z_1)\right)^{\alpha' k_1\cdot k_j}\prod_{j=m+1}^n\vartheta_2(iy_j{-}z_1)^{\alpha' k_1\cdot k_j}.
\end{equation}
The field theory limit of the physical $\I$ terms in the relations is known and given by the Bern-Kosower rules, as explained in the previous subsection. In the next section, we will provide an analysis of the field theory limit of the $\mathcal{J}$ terms in the monodromy relations.

\section{Field theory limit of bulk contours at one-loop}
\label{sec:field-theory-limit}

\subsection{Overall phases}
We analyze here the field theory limit of the $\mathcal{J}$ cycles from~\eqref{string_non_planar}. The relevant part is the piece of the contour running along the $A$-cycle, 
\begin{align}
  \label{eq:bulk-integrals}
   \J_{\mathbf{a}_\pm} :=&\int_0^{\pm 1/2} T_1(x+it) \,\varphi(x+it)\, dx,\\
   \J_{\mathbf{c}_\pm} :=&\int_0^{\pm1/2} T_1(x)\, \varphi(x)\, dx,
\end{align}
with $T_1$ defined as \eqref{T1}. For now we take the string integrand $\varphi=1$ in order to analyze the overall complex phase that these contributions have in the field theory limit. 
For definiteness we look at the integral $\J_{\mathbf{a}_+}$ in detail, while the other phases can be obtained from similar computations.

Define $\mathcal{E}$ as the exponent of the function $T_1=e^{\alpha' \mathcal{E}(z)}$, that is
\begin{equation}
 \mathcal{E}(z)= 2i\pi \ell \cdot k_1 z + \sum_{j=2}^m k_1\cdot k_j
  \ln(-i\vartheta_1(iy_j-z) )+ \sum_{j=m+1}^n k_1\cdot k_j \ln\vartheta_2(iy_j-z).
\end{equation}
Now we take the tropical limit using eq.~\eqref{eq:trop-scal}, but keep the stringy lower-case variables so as to not clutter the notation with factors of $\pi$ and $\alpha'$.
As $\alpha'\to 0$ we can safely drop the higher-order terms in $q$ and $w_j=\exp(2i\pi z_j)$ in
$\ln(-i\vartheta_{1})$ and $\ln(\vartheta_2)$. What remains are logarithms of trigonometric functions, $\ln(-i\sin(\cdot ))$ and $\ln(\cos(\cdot))$ (see appendix \ref{sec:conventions}). The sine and cosine functions are a sum of two terms, one of
which is exponentially growing, the other suppressed. On the $\mathbf{a}_+$ contour, we have $z=x+it$, which gives
\begin{align}
  -i\sin(\pi(i y_j-z)) &= -\frac{1}{2}
                    e^{2\pi(t-ix-y_j)}(1-e^{-2\pi(t-ix-y_j)}),\label{eq:first_phase}\\
  -i \sin(\pi(it-z)) &= i \sin (\pi x),\label{eq:weird_phase}\\
  \cos(\pi(i y_j-z)) &=\frac{1}{2}
                    e^{2\pi(t-ix-y_j)}(1+e^{-2\pi(t-ix-y_j)})
\end{align}
for $j\neq m$.
Upon taking the logarithm we can also discard the exponentially
suppressed terms $e^{-2\pi(t-ix-y_j)}$, which correspond the exchange of massive string states. The exponent then reduces to
\begin{align}
 \mathcal{E}(it+x)\rightarrow \;&2i\pi \ell \cdot k_1 (it+x) + \pi\sum_{j=2}^{m-1} k_1\cdot k_j
 (t-y_j -ix+i\pi)\nonumber\\
 &+\pi \sum_{j=m+1}^n k_1\cdot k_j (t-y_j -ix) +
 k_1\cdot k_m (\sin(\pi x)+i \pi/2).
 \label{eq:exotic-phase-energy}
\end{align}
Here we recover the $i\pi \sum_{j=2}^{m-1} k_1\cdot k_j$ from the term $\ln(-1)=\ln(e^{i\pi})$ in \eqref{eq:first_phase}. This is the usual phase in the monodromy relations obtained from the projection onto the physical branch of the analytically continued function $T_1(z)$.
But note that we have also obtained a term $i \pi k_1\cdot k_m/2$ with an unusual factor of one half, originating from the overall factor of $i=e^{i\pi/2}$ in \eqref{eq:weird_phase}. We can take it to define a canonical branch for these bulk contours. It would be very interesting to see what is its significance in the worldsheet CFT but we leave such considerations to future investigations.

In summary, we find that, in the field theory limit, the contour \eqref{eq:bulk-integrals}, has the following phase:
\begin{equation}
  \label{eq:mono-half}
 \boxed{ \mathlarger{e^{i\pi\alpha' (\sum_{j=2}^{m-1} k_1\cdot k_j + k_1\cdot k_m/2)}}}\,.
\end{equation}
The computation on the $\mathbf{c}_+$ contour is essentially the same and produces the phase
\begin{equation}
  \label{eq:mono-half-c}
   \boxed{ \mathlarger{e^{i\pi \alpha' k_1\cdot k_m/2}}}\,.
\end{equation}
Before concluding this part, let us emphasize the following point. At a generic value of $\alpha'$, the integrals $\J_{\mathbf{a}/\mathbf{c}_\pm}$ are complex, unlike the integrals on the physical contours, which are purely real for real kinematics. When applying monodromies, those physical integrals acquire a phase, which is unambiguous. The situation is different for the integrals $\J_{\mathbf{a}/\mathbf{c}_\pm}$, since even when factoring out the phase mentioned above the integral remains complex.\footnote{These points were already discussed in \cite{Casali:2019ihm} where it was observed that no canonical choice of branch exist for these integrals.}
What the phase above mean is that, taking the field theory limit of the integrands, the first term of the $\alpha'$ expansion is real and given by Feynman graphs, but higher order terms remain complex. Therefore, when we write
\begin{equation}
\J_{\mathbf{a}/\mathbf{c}_\pm}\underset{\alpha'\to 0}{\simeq} e^{i\pi\alpha' (\sum_{j=2}^{m-1} k_1\cdot k_j + k_1\cdot k_m/2)} \times (\mathrm{Feynman\ graphs}+O(\alpha'))\,,\label{eq:J-limit-schem}
\end{equation}
we do not mean that the higher order terms in $\alpha'$ are real, in contrast with the physical integrals where it is indeed true.

\subsection{Contact terms and triangles}
\label{sec:contact_triangle}
We now turn to the evaluation of the integral over $x$ in \eqref{eq:bulk-integrals}. As in the usual Bern-Kosower rules reviewed in section~\ref{subsec:ftl}, there are two cases of interest: when the integrand contains a monomial with exactly one derivative of the worldsheet propagator; and when it doesn't. Taking this into consideration we again write a generic integrand along the $A$-cycles as
\begin{equation}
 \varphi(z_1)=\varphi_1 \dot{G}(z_m-z_1) +\varphi_2
\end{equation}
where $\varphi_1$ and $\varphi_2$ don't contain any monomials on derivatives of the propagator with arguments involving $z_1$. We take the ordering of $\mathcal{J}_{\mathbf{a}/\mathbf{c}_\pm}$ the same as in~\eqref{string_non_planar}, but note that since these calculations are only sensitive to the local structure of the integrand near $z_1$ the results below are valid for any other permutations provided the gauge $z_m=it$ is kept fixed.

\textbf{Triangles:} Triangle graphs are generated by monomials that have exactly one derivative of the worldsheet propagator, which in the field theory limit reduces to a cotangent function, see eq. \eqref{eq:G-dot-def}.
Just like in Bern-Kosower rules, the field theory limit is only affected by local behavior, so it is sufficient to consider the case of $\tilde{\J}_{\mathbf{a}_\pm}$ with $\varphi(x+it)=\dot{G}(z_m-z_1)=\dot{G}(-x)$, that is $\varphi_1=1$ and $\varphi_2=0$; the result will be similar for the contours  $\tilde{\J}_{\mathbf{c}_\pm}$. In the tropical limit, this integral descends to an integral of the form
\begin{equation}
  \label{eq:inttr}
 \J_{\text{trop},\mathbf{a}_\pm} = \int_0^{\pm1/2} \cot(-\pi x) e^{\alpha' i c x + \alpha' k_1\cdot k_m \ln\sin(\pi x)} dx.
\end{equation}
Where momentum conservation was used to rewrite the $x$-dependence of the exponent \eqref{eq:exotic-phase-energy} as
\begin{equation}
\label{eq:f-def}
i c x + k_1\cdot k_m \sin(\pi x) \qquad \text{with}\qquad c=\pi(2\ell +k_m)\cdot
k_1\,.
\end{equation}
This integral can be computed by expanding the exponential $e^{\alpha' i c x}$ since this term is regular when $x\to0$, and we are interested only in the first few orders in $\alpha'$. These give the integrals:
\begin{equation}
  \label{eq:inttr2}
 \J_{\text{trop},\mathbf{a}_\pm}^0=  -\int_0^{\pm1/2} \cot(\pi x) e^{\alpha' k_1\cdot k_m \ln(\sin(\pm\pi
    x))} dx =-\frac{1}{\alpha'\pi k_1\cdot k_m}+{\cal O}(1)
\end{equation}
\begin{equation}
  \begin{aligned}
    \label{eq:inttrcorr}
 \J_{\text{trop},\mathbf{a}_\pm}^1 &= -\alpha' i c x \int_0^{\pm 1/2} x \cot(\pi x) e^{\alpha'
      k_1\cdot k_m \ln(\sin(\pm \pi x))} dx 
    =-\alpha' i c x\frac{\ln(2)}{2\pi}+{\cal O}(\alpha')
  \end{aligned}
\end{equation}
The first integral \eqref{eq:inttr2} produces a tree-like contribution, which is the analogue of the standard case described by the Bern-Kosower rules when two punctures collide on the boundary, except that now it happens in the bulk.\footnote{Note that, contrary to computation in eq.~\eqref{eq:tr-usual}, it was not necessary to zoom in a neighborhood of $z_m=it$, the triangle contribution came from the full integral between $0$ and $\pm 1/2$. This is the case because we already dropped higher order terms, which are exponentially suppressed. Colliding the punctures produces the same pole and cut-off of order $\alpha'$ relative to the pole, which drops out, as in eq. \eqref{eq:tr-usual} and eq. \eqref{eq:tr-usual_2} below.}

The next term in the $\alpha'$ expansion \eqref{eq:inttrcorr} produces terms at two orders higher in $\alpha'$ than the contribution from \eqref{eq:inttr2}, so $ \J_{\text{trop},\mathbf{a}_\pm}^1$ does not contribute in the field theory limit. The second order in $\alpha'$ is important here, because the monodromies can be seen as ${\cal O}(1)$ and ${\cal O}(\alpha')$ relations, therefore a term at first order could possibly enter the first order monodromies. However, we shown below that this does not happen.

In order to fix normalizations we reproduce here also the triangle computation arising from the \textit{physical} contour, where $z_1$ approaches
$z_m=it$, from below, $z_1 \sim i y$:
\begin{equation}
  \label{eq:tr-usual_2}
\begin{aligned}
  \I_{\text{trop}} &= i \int_{t-L}^t \cot(i\pi(t-y)) e^{\alpha' k_1\cdot
    k_m\ln(-i\sin(i\pi(t-y)))}dy \\ &= \frac{(\sinh (\pi L )) ^{\alpha
      k_1{\cdot} k_m}}{\pi \alpha k_1\cdot k_m} = \frac{1}{\alpha' \pi
    k_1\cdot k_m}+\frac{\log (\sinh (\pi L))}{\pi }+O(\alpha').
\end{aligned}
\end{equation}
and we see that triangles from the $A$-cycle contours and from the physical contours have the same normalization. As explained in \eqref{eq:tr-usual}, $L$ is an IR cut-off which drops off when taking into account the part of the integration that is between $y_{m-1}$ and $t-L$.

\begin{figure}
	\centering
 \includegraphics[scale=0.9]{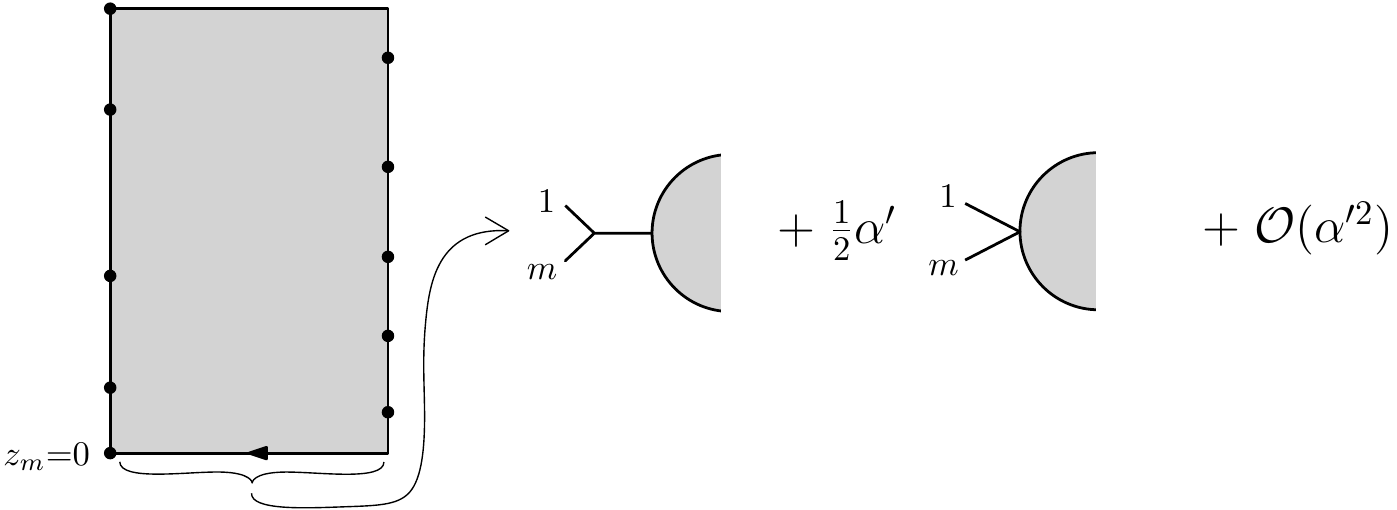}
 \caption{Graphs appearing in the field theory limit of the bulk cycles.}
 \label{fig:bulk_contributions}
\end{figure}

\textbf{Contact term:} Finally, we return to the bulk contour and investigate the case $\varphi=\varphi_2=1$, so that it does not contain a derivative of the Green's function and no subtree is generated. The integration can be done explicitly and leads to
\begin{equation}
  \label{eq:no-tr}
  \int_0^{\pm1/2}e^{\alpha'k_1\cdot k_m \ln \sin(\pm \pi x)}dx =\frac{\Gamma
    \left(\frac{1}{2} (\alpha' k_1\cdot k_m +1)\right)}{2 \sqrt{\pi }
    \Gamma \left(\frac{\alpha' k_1\cdot k_m}{2}+1\right)}=\pm\frac12
  \mp\alpha'k_1{\cdot} k_m{\ln(2)}/2+O(\alpha'^2)
\end{equation}
giving an important factor of $\pm1/2$ at leading order. Since in the limit the particles $z_1$ and $z_m$ are squeezed together, we interpret the leading order in \eqref{eq:no-tr} as a diagram with a contact term involving particles $1$ and $m$ as depicted in figure~\ref{fig:bulk_contributions}.

Note that since we do not rescale $z_1$ in the field theory limit the contributions from the above integrals come at a higher order in $\alpha'$ than the other terms in the monodromy relations. This is easily seen by looking at how the moduli space measure scales for different terms in the monodromy relations~\eqref{string_non_planar}. In the $\mathcal{I}$ contributions all coordinates are rescaled in the field theory limit such that
\begin{equation}\label{eq:I_meas}
 d \tau\prod_{i=1;i\neq m}^{n}d z_i\rightarrow (\alpha')^{-n} dT\prod_{i=1;i\neq m}dY_i
\end{equation}
while the contributions from $\mathcal{J}$ have an overall $(\alpha')^{n-1}$ from the measure since the coordinate $z_1$ is not scaled in the limit. Another way of seeing this is to realize that each propagator in a trivalent graph in the field theory limit has to come with a factor of $\pi\alpha'$. This comes from the fact that propagators are generated in the tropical limit by integrals of the tropical Koba-Nielsen factor, where Mandelstam variables are multiplied by $\pi\alpha'$. This implies that the contact term comes with an overall extra $\pi\alpha'$ term compared with the other terms.

Note also the presence of a relative factor of $-i$ in front of the contact term coming from the measure: the original integration contour for $z$ along the imaginary axis is such that under analytic continuation $dy=-i dz$. Since the $\mathcal{J}$ terms are integrated along the real axis where $dz = dx$, a relative factor of $i$ should be added. The triangle terms originating from $\mathcal{J}$ also have this extra $-i$ for the same reason but due to another factor of $i$ coming from the analytic continued $-i\vartheta_1(z_m-z_1)$ in the integrand the overall coefficient of these terms is $-1$, the same as if particle $1$ was above $m$ in a contribution away from the edges. 

\section{Field theory limit of the monodromy relations}
\label{sec:field-theory-limit-monodromy}

Let us now turn to studying the field theory limit of the relations.
We start by reviewing the tree-level setup to recall how the limit works in this simple case.

At finite $\alpha'$, the monodromy relations~\eqref{string_non_planar} are two linearly-independent identities. Plahte's relations~\eqref{eq:plahte}, can therefore be rewritten as their sum and differences, leading to
\begin{equation}
  \begin{aligned}
  \sum_{i=1}^{n-1}\cos(\pi \alpha' k_1\cdot \sum_{j=2}^ik_j)A_{\text{tree}}(2,\dots, i, 1, i{+}1,\dots,n ) =0,\\  \sum_{i=1}^{n-1}\sin(\pi \alpha' k_1\cdot \sum_{j=2}^ik_j)A_{\text{tree}}(2,\dots, i, 1, i{+}1,\dots,n ) =0.
\end{aligned}
  \label{eq:plahte-trigo}
\end{equation}
As $\alpha'\to0$, the trigonometric functions are simplified as $\sin(\pi\alpha' s)\simeq \pi\alpha' s$, $\cos(\pi\alpha' s)\simeq 1$ and $A_{\text{tree}}$ descends to the field theory amplitude $A_{\text{tree}}^{\text{FT}}$, leading to
\begin{align}
  \sum_{i=1}^{n-1}A_{\text{tree}}^{\text{FT}}(2,\dots, i, 1, i{+}1,\dots,n ) =0,   \label{eq:KK}\\
    \sum_{i=1}^{n-1} (k_1\cdot \sum_{j=2}^ik_j) A_{\text{tree}}^{\text{FT}}(2,\dots, i, 1, i{+}1,\dots,n ) =0.\label{eq:fundbcj}
\end{align}
The first relations are the Kleiss-Kuijf relations~\cite{Kleiss:1988ne}, while the second are the fundamental BCJ relations, known to be equivalent to BCJ relations between graphs \cite{Feng:2010my,Feng:2011fja}, as initially observed in \cite{BjerrumBohr:2009rd,Stieberger:2009hq}.

A more cavalier way to arrive at these identities is by taking the field theory limit before taking their linear combinations in \eqref{eq:plahte-trigo}, and extract the ${\cal O}(1)$ and ${\cal O}(\alpha')$ from the phases only. This is requires, in theory, to check that no $O(\alpha')$ arise from the amplitude themselves and mixes up with the $\alpha'$ coming form the amplitudes. Both relations obtained in both ways are identical.\footnote{An interesting related point, which, to our knowledge, was not raised in the literature before, is the following (see also \cite{Broedel:2012rc}). From the $+$ relation, involving only cosines of the phases, one can see that any order $\alpha'$ correction to the amplitude $A^{FT}$, $A_{\alpha'}$, must satisfy independently the Kleiss-Kuijf relations of the form $\sum_\mathrm{permuations}A_{\alpha'}=0$.
Therefore, any ${\cal O}(\alpha')$ corrections to the amplitude which would occur in the ${\cal O}(\alpha')$ relations would cancel up independently of the ${\cal O}(\alpha')$ terms coming expanding the phases. Indeed, the ${\cal O}(\alpha')$ relations would write $\sum \alpha' k\cdot k A^{FT}+ \sum A_{\alpha'}=0$ and the second sum vanishes, in virtue of what was said above.}

At loop level, it is not hard to convince oneself that this shortcut reasoning works for the standard part of the relations, which involve usual $\mathcal{I}$ integrands, since they are purely real. But there are also new terms, the $\mathcal{J}$ bulk contours, which are complex and do not have a well-defined phase. Therefore terms of the form $\J_{+}\pm \J_-$ occur and one needs to be careful in handling them. It turns out that the analysis at orders ${\cal O}(1)$ and ${\cal O}(\alpha')$ continues to hold, due to (anti)-symmetry properties of the $\J_{\pm}$'s under complex conjugation. Therefore we shall present the analysis of the field theory limit in terms of ${\cal O}(1)$ and ${\cal O}(\alpha')$ relations directly, rather than sums and differences, as it is less cumbersome.

Following up on the discussion above, we consider only the $+$ monodromy relation in eq.~\eqref{string_non_planar}.
They take the schematic form:
\begin{equation}
  \sum_i e^{i\alpha' \pi \phi_i} \mathcal{I}_i+\sum_j e^{i\alpha' \pi  \theta_j}\mathcal{J}_j=0\label{eq:mono-schem}\,,
\end{equation}
where the phases $\phi_j$ and $\theta_j$ consist of combinations of external and internal kinematic invariants, and the indices $i,j$ runs over the set of contours in the relation. In the field theory limit, our analysis in section \ref{sec:field-theory-limit} shows that the integrands behave as
\begin{align}
\mathcal{I}_i &=\mathcal{I}^{\text{FT}}_i +\mathcal{O}(\alpha'^2),\\
\mathcal{J}_j &=-\mathcal{J}^{\text{FT}}_j -i\pi\alpha'\mathcal{J}^{\text{CT}}_j +\mathcal{O}(\alpha'^2),
\end{align}
where the superscript $\text{FT}$ denotes contributions from the usual trivalent graphs appearing in the field theory limit as described in the previous section. In other words, $\I^\mathrm{FT}$ and $\J^\mathrm{FT}$ are simply sums of trivalent graphs with a common definition of the loop momentum. The superscript $\text{CT}$ denotes contact terms that appear with an extra factor of $-i\pi\alpha'$ in front as described at the end of section~\ref{sec:field-theory-limit}.
Other signs and factors of $i\pi \alpha'$ are also carefully explained there. In other words, $\J^\mathrm{CT}$ is the result of taking all the graphs appearing in  $\J^\mathrm{FT}$ and removing the legs generated by subtrees from the bulk integrals. 

In the field theory limit the monodromy relations are therefore written as
\begin{equation}
  \label{eq:mono-FTL-schem}
  \left( \sum_i \mathcal{I}^{\text{FT}}_i-\sum_j\mathcal{J}^{\text{FT}}_j \right)+ i\pi\alpha'\left(  \sum_i \phi_i \mathcal{I}^{\text{FT}}_i - \sum_j\theta_j\mathcal{J}^{\text{FT}}_j- \mathcal{J}^{\text{CT}}_j\right) + {\cal O}(\alpha'^2)=0\,.
\end{equation}
As noted above, and originally observed in~\cite{Tourkine:2016bak}, these relations split into two different sets. The ${\cal O}(1)$ term constitute the Bern-Dixon-Dunbar-Kosower relations \cite{Bern:1994zx,DelDuca:1999rs} at one loop, and contain the relations found in \cite{Feng:2011fja} at two loops as shown in \cite{Tourkine:2016bak}. For the planar case at ${\cal O}(1)$ they are the Boels-Isermann relations~\cite{Boels:2011mn,Boels:2011tp}.

Our goal in the next two subsections is to study in detail these two relations, and in particular characterize exactly the types of graphs which enter the relations.

\subsection{Cancellations at $\mathcal{O}(1)$}
\label{sec:cancelations-at-o1}

\begin{figure}
\centering
 \includegraphics{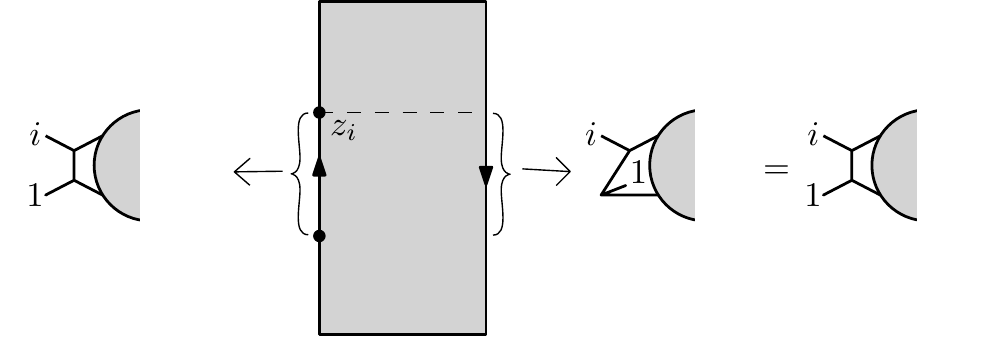}
 \caption{Contributions from different boundaries that cancel each other.}
\label{fig:box_cancel}
\end{figure}

At $O(1)$, there are two main classes of graphs that we need to investigate: those coming from an integrand containing a monomial on $\dot{G}_{i1}$, and the others. As explained in section \ref{sec:field-theory-limit}, the former produces a tree where particles $i$ and $1$ attach to the rest of the graph; the latter produces a propagator connecting particles $i$ and $1$.\footnote{Contact terms can also be produced in the $\mathcal{J}$ contours but these only contribute at higher order in $\alpha'$.} These two situations can be visualized as two ways of attaching particles $i$ and $1$ to the main graph using a ``bridge'', as in figure~\ref{fig:box_cancel}.
The cancellation within these classes of diagrams is immediate: for each diagram with a propagator between $1$ and $i$ originating from a contour where $z_1$ is in one boundary, the same diagram is also generated from a term where $z_1$ is in the other boundary but with a negative sign, see figure~\ref{fig:box_cancel}. Since at this order the phases in the monodromy relation do not contribute, these diagrams cancel term by term.

\begin{figure}
\centering
 \includegraphics{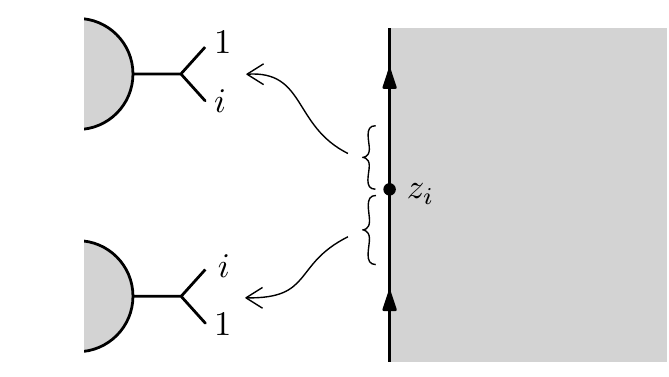}
 \caption{Graphs with attached trees that cancel each other due to a relative minus sign.}
\label{fig:triangle_cancel}
\end{figure}

The diagrams with attached trees also cancel term by term, but care must be taken for the edge cases, that is, when $z_1$ approaches $0$ or $it$. If the triangle diagrams originate from a contour away from $z=0$ or $z=it$ as in figure~\ref{fig:triangle_cancel}, they cancel term by term due to the antisymmetry of $\dot{G}_{i1}$ in the integrand. This is explicitly shown below. In \eqref{eq:tr-usual-1}, $z_1$ approaches $z_i$ from below, and in \eqref{eq:tr-usual-2}, where approaches $z_i$ from above: 
\begin{multline}
  \label{eq:tr-usual-1}
  \I_{\text{trop}}^0 = i \int_{y_i-L}^{y_i}\cot(i\pi(y_i-y)) e^{\alpha' k_1\cdot
    k_m\ln(-i\sin(i\pi(y_i-y)))} dy\\ = \frac{(\sinh (\pi L )) ^{\alpha'
      k_1\cdot k_i}}{\pi \alpha' k_1\cdot k_i} = \frac{1}{\alpha' \pi
    k_1\cdot k_i}+\frac{\log (\sinh (\pi L))}{\pi }+ {\cal O}(\alpha'),
\end{multline}
\begin{multline}
  \label{eq:tr-usual-2}
  \I_{\text{trop}}^1 = i \int_{y_i}^{y_i+L} \cot(i\pi(y_i-y)) e^{\alpha' k_1\cdot
    k_m\ln(-i\sin(i\pi(y-y_i)))}dy \\=- \frac{(\sinh (\pi L )) ^{\alpha'
      k_1\cdot k_i}}{\pi \alpha' k_1\cdot k_i} =- \frac{1}{\alpha' \pi
    k_1\cdot k_i}-\frac{\log (\sinh (\pi L))}{\pi }+ {\cal O}(\alpha').
\end{multline}

Since phases do not contribute at this order, these diagrams are simply summed and therefore cancel. Note that the dependence on the $L$ is pushed to order $(\alpha')^2$, preventing the risk of spoiling the $O(\alpha')$ relations. As argued in section~\ref{subsec:ftl}, all dependence in $L$ should vanish in the full contour.

At the edges, $z_1=0$ or $z_1=it$, the argument given above does not apply. There are triangles generated by these edge terms, see figure~\ref{fig:edge_triangle} a) and b), but they come with different labeling of the loop momenta and cancellation is not possible at fixed loop momentum with just these terms.
However, as demonstrated in section~\ref{sec:contact_triangle}, the $\mathcal{J}$ terms produce diagrams with trees attached with the right propagator prefactor to cancel these edge contributions.
The integrals in eqs.~\eqref{eq:inttr} and \eqref{eq:tr-usual_2} give  precisely the correct sign for this cancellation to occur. Moreover, they have the same loop momentum assignment, which gives a complete cancellation, at fixed loop momentum. These extra contributions are depicted in figure~\ref{fig:edge_triangle}, c) and d).

This concludes our analysis of the field theory limit of the monodromy relations at order ${\cal O}(1)$.
\afterpage{\begin{figure}
  \centering
  \includegraphics{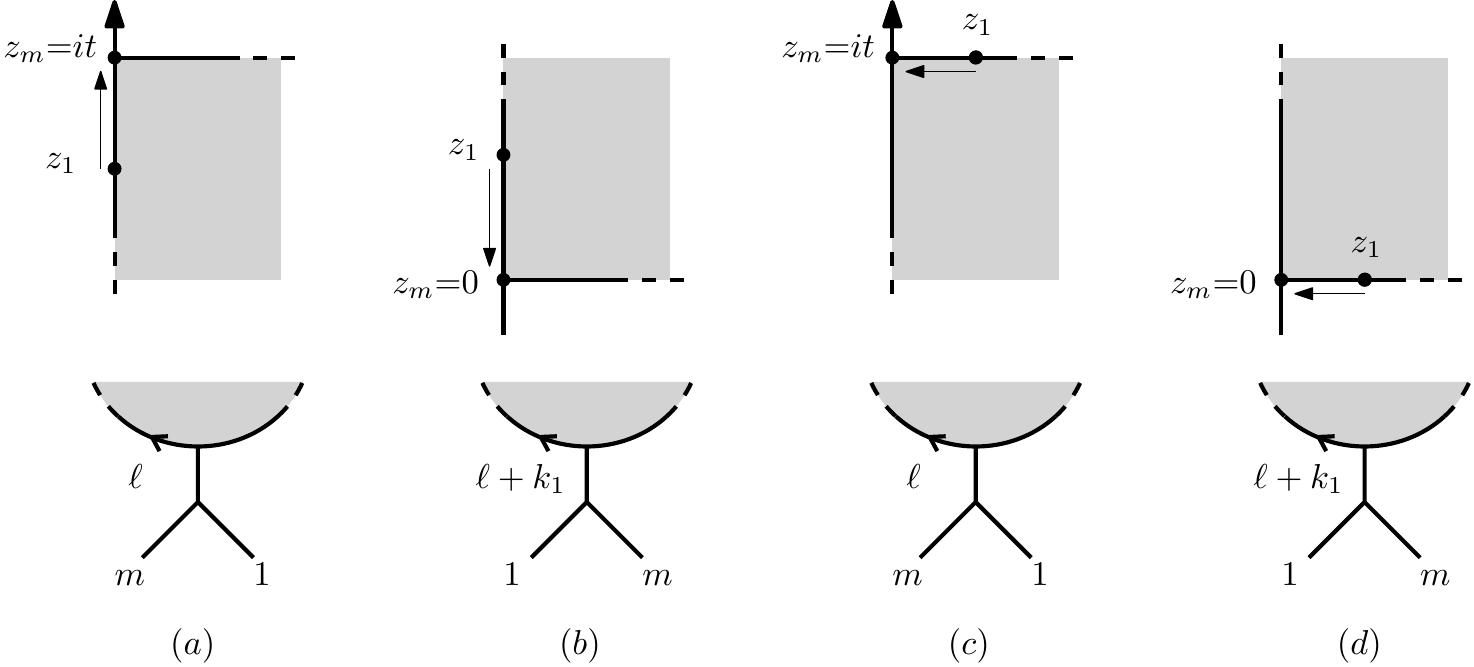}
  \caption{Four graph degenerations near the worldsheet edges, $(a$-$d)$. \underline{Top:} Worldsheet (gray) with puncture $z_1$ approaching $z_m$ from various boundaries. \underline{Bottom:} Corresponding graphs with resulting $(1m)$ tree pinching off.}
  \label{fig:edge_triangle}
\end{figure}}

\subsection{Cancellation at $\mathcal{O}(\alpha')$ and BCJ triples}
\label{sec:canc-at-mathc}

The only new graph that has to be considered is the contact term that appears at $\mathcal{O}(\alpha')$ in the expansion of the $\mathcal{J}$'s, see figure~\ref{fig:bulk_contributions}. This is a structurally new mechanism. Unlike before, where at ${\cal O}(\alpha')$ only the phases from the exponentials contributed, here we see an order ${\cal O}(\alpha')$ coming from the diagram itself. It is remarkable to see that this term precisely cancels the other graphs generated by the monodromies. Notwithstanding, the cancellation itself is not surprising since we know that the relations hold at finite $\alpha'$ \cite{Casali:2019ihm} and the field theory limit is simply a consequence of these relations.

Like at tree-level, the cancellation will depend on two mechanisms: the fact that the $\mathcal{O}(\alpha')$ terms of the phases conspire to cancel propagators in the graphs resulting in the grouping of numerators from different graphs into BCJ triples; and that the string theory, through the Bern-Kosower rules, produces automatically color-kinematics satisfying numerators, in the field theory limit, for the graphs where the definition loop momentum does not jump in a Jacobi identity.
To our knowledge, this last observation is new and we shall devote some time explaining it now.

\paragraph{Numerators away from the $A$-cycle.}
For cycles away from the edges we can argue that numerators obey color-kinematics as follows (see also \cite{Mizera:2019blq}). Consider a Jacobi identity with legs $1$ and $j$. Generically, a string theory numerator will be of the form
\begin{equation}
  \varphi = \dot{G}_{j1} \varphi_1 +\varphi_2,
  \label{eq:num-dec}
\end{equation}
where $\varphi_{1}$ and $\varphi_{2}$ do not contain any terms with single poles as $z_1\rightarrow z_j$ (this definition allows  $\varphi_2$ to have a $\dot{G}_{j1}^2$ term). This integrand gives three types of graphs in the field theory limit, each with its respective numerator $n_s,\,n_t$ or $n_u$, see figure~\ref{fig:stu}.
The term $\dot{G}_{j1}\varphi_1$ produces diagrams with attached trees, which we call \emph{triangle-type}, and diagrams with propagators connecting $1$ and $j$, which we call \emph{box-type}, in analogy with the situation at four points. These come with numerators
\begin{equation}
  n_{1,\text{triangle}} = 2 \varphi_1,\qquad n_{1,\text{box}} =  \varphi_1,
\end{equation}
where the factor of $2$ comes from rewriting the propagators in the standard form $1/(k_1 \cdot k_j ) = 2/(k_1+k_j)^2$. The term $\varphi_2$ produces only box type graphs with numerator
\begin{equation}
  n_{2,\text{box}} = \varphi_2.
\end{equation}
\afterpage{\begin{figure}
  \centering
  \includegraphics{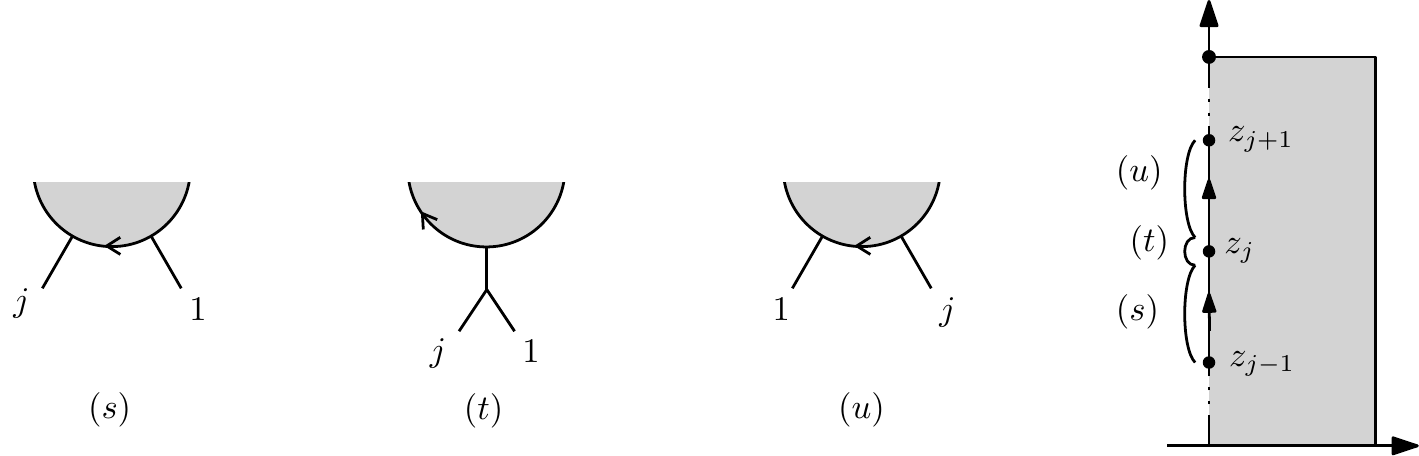}
  \caption{Illustration where the $s$-, $t$-, $u$-channel graphs come from on the worldsheet.}
  \label{fig:stu}
\end{figure}}
The numerators for each graph are the following combinations of the above:
\begin{equation}
  \begin{aligned}
    n_s& = n_{1,\text{box}} + n_{2,\text{box}},\\
    n_t& = n_{1,\text{triangle}} ,\\
    n_u& = -n_{1,\text{box}} + n_{2,\text{box}},
  \end{aligned}
\end{equation}
which can be immediately checked to satisfy
\begin{equation}
 n_s-n_t=n_u.
\end{equation}
This is the kinematic Jacobi identity. We conclude that numerators given by string theory away from the origin of the loop momentum obey color-kinematics duality.

\paragraph{Numerators near the $A$-cycle.}
The above argument cannot hold near the $A$-cycles of the worldsheet since the monodromy relations do not have the required contours either above or below the $A$-cycle where the surface is cut.
However, there are contributions along this cycle, the $\mathcal{J}$ integrands, which exactly cancel the leftover numerators. To see this explicitly consider again a generic integrand of the form \begin{equation}\label{eq:gen_2}
 \varphi=\dot{G}_{m1}\varphi_1+\varphi_2,
\end{equation}
where $z_m$ is the position of the particle fixed at the corner of the cut worldsheet. There are six contours that contribute to graphs where $1$ and $m$ are adjacent, divided in two sets of three coming either from the top or the bottom of the rectangle.%

\afterpage{\begin{figure}
	\centering
	\includegraphics{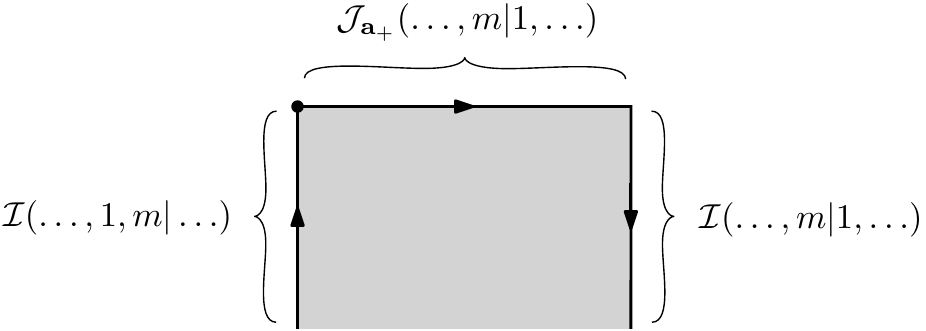}
	\caption{Contours that produce graphs where particles $1$ and $m$ are adjacent in the field theory limit and which have mutually consistent definition of the loop momentum. Three more contours are on the lower side of the annulus.}
	\label{fig:edge_contours}
\end{figure}}

\paragraph{Consistency with monodromy relations near the $A$-cycle.}
For example, one set of diagrams with mutually consistent definition of loop momenta comes from a term where both particles are on the same boundary $\mathcal{I}(\dots 1,m|\dots)$, one where they are on different boundaries $\mathcal{I}(\dots m|1\dots)$, and one where $1$ is integrated along an $A$-cycle: $\mathcal{J}(\dots m|1\dots)$. These are depicted in figure~\ref{fig:edge_contours}. The monodromy phases produce a coefficient of the box-type diagram that can be simplified:
\begin{equation}
 \label{eq:box_coeff}k_1\cdot\sum_{i=2}^{m-1}k_i-k_1\cdot \left(\sum_{i=2}^m k_i-\ell\right)=-k_1\cdot(k_m-\ell)=\frac{(k_m-\ell)^2}{2}-\frac{(k_1+k_m-\ell)^2}{2}.
\end{equation}
The second term cancels the propagator between $1$ and $m$ in these diagrams leaving behind a contact diagram with coefficient $\frac{1}{2}(\varphi_1+\varphi_2)$. Recall that $\mathcal{J}$ produces a contact diagram with exactly the same form as the one leftover from the box type but with a coefficient $-\frac{1}{2}\varphi_2$. This cancels one of the terms leaving only the $\frac{1}{2}\varphi_1$. The last term should be canceled by the triangle-type term, which can be seen easily: both $\mathcal{I}(\dots 1,m|\dots)$ and $\mathcal{J}(\dots m|1\dots)$ produce triangles with phase coefficients
\begin{equation}
 \label{eq:tri_phase}
 k_1\cdot \left(\sum_{i=2}^{m-1}k_i\right)-k_1\cdot\left(\sum_{i=2}^{m-1}k_i+\frac{k_m}{2}\right)=-\frac{1}{2}k_1\cdot k_m,
\end{equation}
which, combined with the integrand numerator $\varphi_1$, cancels the leftover term exactly. In this cancellation we see the crucial role played by the unusual $1/2$ phase that appears in $\mathcal{J}$.
More precisely, we have:
\be
\includegraphics[scale=.9,valign=c]{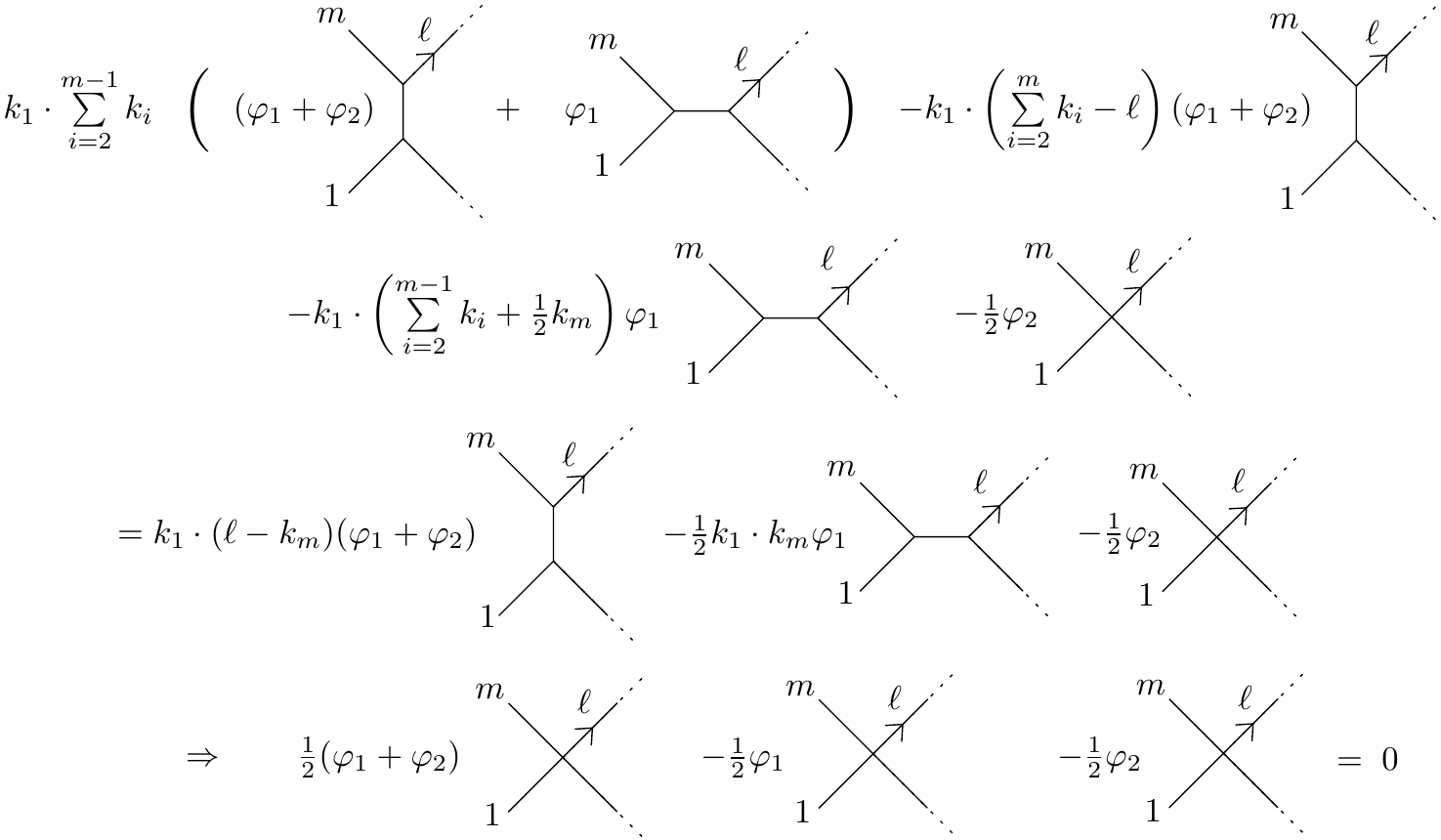}
\ee

\begin{figure}[t]\centering
 \includegraphics{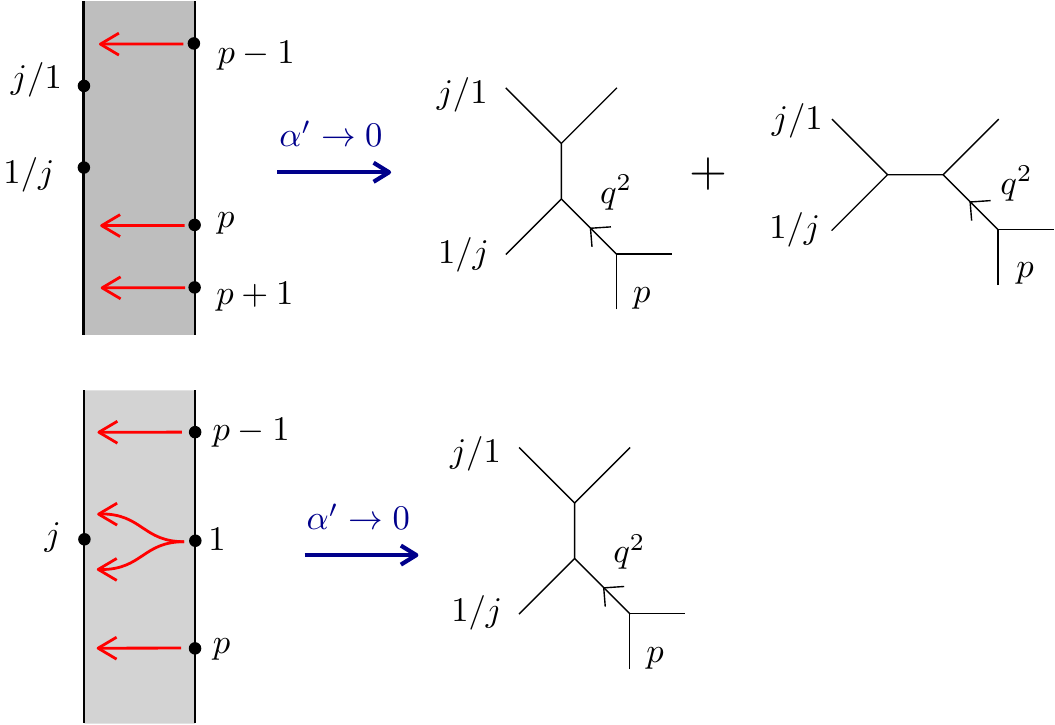}
 \caption{How graphs participating on a particular BCJ triple are generated. \underline{Top:} Planar graphs contribute to tree and box-type graphs. \underline{Bottom:} Non-planar graphs only contributes to box-type diagrams. (We use $q^2=(\sum_{i=2}^{j-1}k_i+\sum_{i=p}^nk_i+\ell)^2 =(\ell-\sum_{i=j}^{p-1}k_i-k_1)^2$.)}
  \label{fig:general_prop}
\end{figure}

\paragraph{Consistency with monodromy relations away from the $A$-cycle.}
As described in \cite{Ochirov:2017jby,Tourkine:2019ukp}, the mechanism of cancellation of the propagators also works for cycles that lie away from the $A$-cycles.
To see this in detail consider the generic non-planar terms $\mathcal{I}(\dots,1,j,\dots|\dots,p,\dots)$ and $\mathcal{I}(\dots,j,1,\dots|\dots,p,\dots)$ where particle $1$ is away from the edge and particle $p$ is some fixed particle with which we use to group together graphs with the same internal momenta as in figure~\ref{fig:general_prop}. The graphs of interest are those where in the field theory limit the worldsheet degenerates such that particle $p$ sits right before $1$ an $j$ in the resulting cyclic ordering.
A tree-type graph can only be generated when $1$ and $j$ sit on the same boundary, so it comes with a coefficient
\begin{equation}
 k_1\cdot\sum\limits_{i=2}^j k_i ,
\end{equation}
which cancels the propagator in the tree-type graph with $1$ right after $j$, giving a contact term with numerator $n_t$. The two box-type graphs are generated from both the planar and non-planar string amplitude, in this case their phase coefficients combine:
\begin{align}\label{eq:phase_1_j}
 k_1\cdot\left(\sum_{i=2}^{j-1}k_i-\sum_{i=2}^{p-1}k_i+\ell\right)&=-k_1\cdot\left(\sum_{i=j}^{p-1}k_i-\ell\right)\nn\\
 &=\frac{1}{2}\left(\ell-\sum_{i=j}^{p-1}k_i\right)^2-\frac{1}{2}\left(\ell-k_1-\sum_{i=j}^{p-1}k_i\right)^2
\end{align}
for the diagram with $1$ before $j$ and
\begin{equation}\label{eq:phase_j_1}
\frac{1}{2}\left(\ell-\sum_{i=j+1}^{p-1}k_i\right)^2-\frac{1}{2}\left(\ell-k_1-\sum_{i=j+1}^{p-1}k_i\right)^2
\end{equation}
for the diagram with $j$ before $1$. These give four canceled propagators but only two of them are related to the BCJ triple where $1$ and $j$ are exchanged, that is, we look for the terms that cancel the propagator between $1$ and $j$. Minding the signs (all particles are taken as incoming), these are the first term in \eqref{eq:phase_1_j} and the second term in \eqref{eq:phase_j_1}, the other two give contributions to other BCJ triples. The result is two  contact terms with numerators $n_s$ and $n_u$. Since we canceled the propagator between $1$ and $j$ for these box-type graphs and the exchange propagator for the tree-type graph, they all produce the same contact term graph with a numerator given by the sum
\begin{equation}
 n_s-n_t-n_u=(\varphi_1+\varphi_2)-2\varphi_1-(-\varphi_1+\varphi_2)=0.
\end{equation}
In more detail, we have:
\be
\includegraphics[scale=1,valign=c]{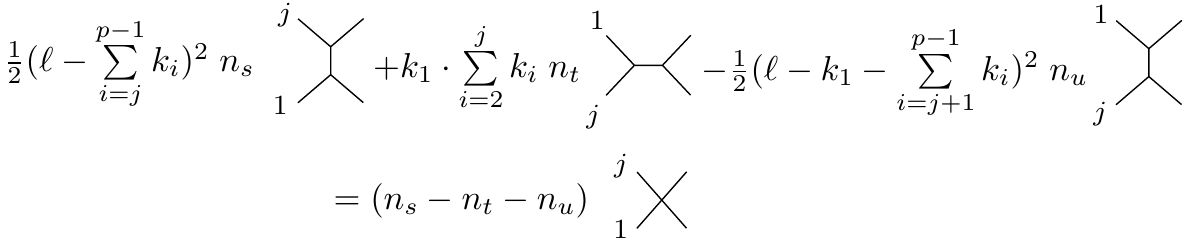}
\ee

This proof is completely generic and leads us to the conclusion that any set of numerators obtained from the Bern-Kosower rules at one loop, after bringing it to the form \eqref{eq:schem}, is automatically in a BCJ satisfying representation. Again, this analysis holds for the numerators of those graphs involved in BCJ identities which do not relabel the loop momentum.

\section{Discussion}
\label{sec:discussion}

\subsection{Summary of results}
\label{sec:summary-results}

\paragraph{Field theory limit of the bulk cycles.}
As we have seen in the text, the bulk contours contribute significantly to both in the $O(1)$ and $O(\alpha')$ relations, at fixed loop momentum.\footnote{It was first observed in~\cite{Tourkine:2016bak,Hohenegger:2017kqy} that the ${\cal O}(1)$ relations can be seen as amplitudes relations once the loop momentum is integrated out, these are the Bern-Dixon-Dunbar-Kosower relations \cite{Bern:1994zx}. As the bulk integrals simply differ by a relabelling of the loop momentum they cancel each other out after integration since the phases don't contribute at this order.} %
However, at fixed loop momentum, as shown above, they are needed for an exact cancellation between graphs (including tree-type graphs) already at $O(1)$, which are identical but for the definition of the loop momentum.

At $O(\alpha')$ they are essential for complete cancellation. We showed that in the field theory limit of these cycles there is an unconventional phase in front of the $\mathcal{J}$ contours. This is a hitherto new phase, not seen before:
\begin{align}
 e^{\pm\frac{1}{2}i\alpha'\pi k_1\cdot k_m}.
\end{align}
The unusual factor of $1/2$ was crucial for a complete cancellation among diagrams at fixed loop momentum. 

As mentioned previously there's no physical justification from string theory as for the existence and importance of those bulk contours, although they are unavoidable in the theory of twisted cycles~\cite{Casali:2019ihm}. Our analysis shows that they are related to ambiguities that arise in shifting loop momentum and restore the well-definedness of the string integrand under loop momentum redefinition.

\paragraph{BCJ numerators in Bern-Kosower representations.} In checking the exact cancellations between all graphs appearing in the monodromy relations, we also studied ``standard'' cancellations, where no ambiguity linked to loop-momentum definition arise. We found that all Bern-Kosower numerators satisfy BCJ identities away from the points where loop-momentum jumps. To our knowledge, this is a new result. Below, we comment on the fact that we expect this to hold to any loop order, since it stems from the elementary properties of the derivative of the worldsheet Green's function; antisymmetry, and local behavior.

\subsection{Towards KLT and higher loops}
\label{sec:towards-klt-loops}

\paragraph{Contact-terms of higher valency in a KLT formulation.}

In \cite{Casali:2019ihm}, we argued that a complete basis of integration cycles for the Kawai-Lewellen-Tye \cite{Kawai:1985xq} construction at loop-level has to include bulk cycles. The bulk cycles described there there, and used here, are those where only one particle moves in the interior the annulus. Because the construction~\cite{Casali:2019ihm} was recursive, it is not hard to see that twisted cycles have to include cycles where more than one particle is inside the annulus.\footnote{Actually, apart from the one particle whose position is gauge fixed by translation invariance, all particles could even be on those cycles. Which of those cycles would be related by monodromy relations can also be worked out within our formalism.}

In the tropical limit, those higher-bulk cycles would generalize our analysis, and give rise to contact terms where $k$ propagators are canceled, if $k$ particles sit in the bulk. A more thorough analysis is left over for future work.

Following up on the discussion in~\cite{Casali:2019ihm}, one could therefore speculate about the form of a KLT formula at loop level. Since the bulk cycles are unavoidable, and since in the field theory limit they create contact-terms, we are lead to expect the following. In a general formulation of the double copy in string theory, \'a la KLT, one should expect to find squared contact-term graphs and products of ordinary graphs and contact-term graphs.

This would arise from a KLT formulation, where twisted cycles are multiplied together by a yet-to-be discovered momentum kernel. 

It is however interesting to remark that, in the context of the generalized double copy \cite{Bern:2017yxu,Bern:2017ucb}, contact-terms were only needed at high loop orders (five loops), while they would seem to appear naturally already at one-loop.

This conundrum makes the question of figuring out a KLT formula at loop level even more pressing and interesting.

\paragraph{Higher-loop generalizations.} While extension of our results to higher-loop integrands requires extra care and is left for future work, let us briefly comment on which parts generalize straightforwardly and which need new analysis.

\begin{figure}
	\centering
	\includegraphics{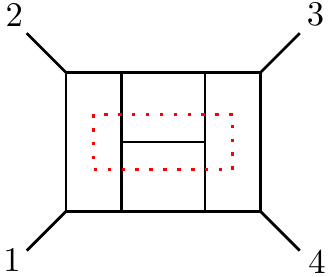}
	\caption{Our Jacobi identities do not involve doing purely internal BCJ moves at higher loops.}
	\label{fig:internal-triple}
\end{figure}

First of all, let us delineate which BCJ triples we expect to be describable within our framework. Higher-genus monodromy relations, at least as described in \cite{Tourkine:2016bak,Hohenegger:2017kqy,Casali:2019ihm}, only involve identities corresponding to a puncture traveling along boundaries of the cut Riemann surface. Therefore, these identities always involve BCJ triples that have at least one external leg. An example of an ``internal'' BCJ triple at four loops, which is not contained in our framework, is shown in figure~\ref{fig:internal-triple}.

Secondly, the Jacobi identities related to triplets without loop-momentum ambiguity (away from the $A$-cycles) have been described in \cite{Tourkine:2019ukp,Casali:2019ihm}. The phase factors of the monodromies were shown to adequately cancel neighboring propagators and regroup graphs by BCJ triplets. Indeed, one can straightforwardly extend the reasoning of section~\ref{sec:canc-at-mathc} and show that, if a Bern-Kosower representation is provided at higher loops for a worldline integrand, it has to satisfy color-kinematics duality. In other words, granted that Bern-Kosower representations may be found at higher loops, we claim that corresponding numerators which do not redefine the origin of loop momenta should obey Jacobi identities. \footnote{Concerning this latter question, one may worry about a few things. In the RNS formalism, the supermoduli space non-splitness~\cite{Donagi:2013dua,Witten:2012bh} prevents the naive existence of a purely bosonic integrand, which is the type of the Bern-Kosower representations. Sen's vertical integration~\cite{Sen:2014pia, Sen:2015hia} is in principle a prescription to project the supermoduli space integral to a bosonic one, but nothing is known about whether or not this could lead to Bern-Kosower representations. One may for instance worry that during the integration by parts process to reach a Bern-Kosower integrand (with single derivatives of the Green's function), vertex operators could collide with picture changing operators. If this happens, additional contact-term like contributions may be expected. In purely bosonic representations (bosonic string, pure spinors), there are no such obstructions and, since the integration by part sequence which leads to a given Bern-Kosower integrand does not depend on the genus of the Riemann surface, one may expect these representations to exist.}

\begin{figure}
	\centering
	\includegraphics[scale=0.8]{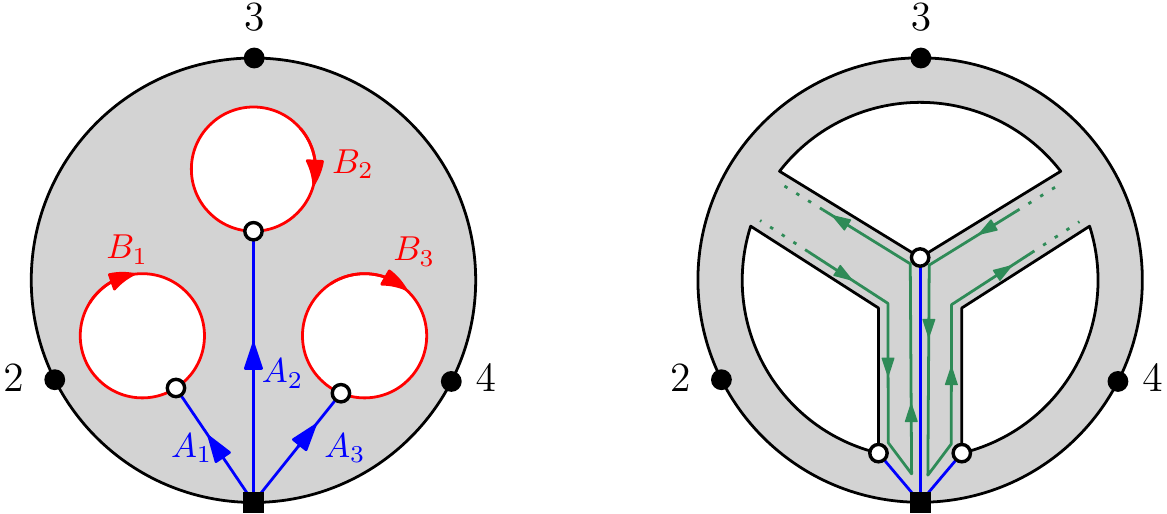}
	\caption{Mercedes graph degeneration. \underline{Left:} The decomposition of the Riemann surface in terms of $A$- and $B$-cycles. The loop momenta $\ell_I$ are flowing through $A_I$ for $I=1,2,3$. \underline{Right:} The $\J$ cycles (green) get trapped between the $B$ cycles in the neighborhood of the Mercedes degeneration.}
	\label{fig:J-pinch}
\end{figure}

Finally, we can discuss triples in the neighborhood of the $A$-cycles: those suffer from the loop-momenta ambiguities analogous to the ones at one loop. Here a genuinely new feature arises: the $\J$ cycles can be pinched between holes of a Riemann surface in the tropical limit, cf. figure~\ref{fig:J-pinch} for a 3-loop Mercedes diagram degeneration. Apart from this important issue, which will be carefully addressed elsewhere, the analysis for other types of degenerations is quite similar to the one at genus one.

Let us therefore finish the paper with a brief discussion of those cases and illustrate it with a two-loop example.

The arguments of section~\ref{sec:canc-at-mathc} generalize to higher loops immediately, as they depend only on the local properties of derivatives of the Green's function: its antisymmetry on the worldline $\partial_{z_i} G^{WL}(z_i,z_j) = -\partial_{z_i} G^{WL}(z_j,z_i)$ and the local behavior $\partial_z G(z) \sim 1/z$, which gives rise to subtree graphs required for $t$-channel graphs.\footnote{See \cite{Dai:2006vj,Tourkine:2013rda} for the explicit expressions of the worldline Green's function at higher genus coming from the tropical or field theory limit of string theory. Along an edge, it is given by the geometric distance between two points, whose derivative is just the sign function, and therefore antisymmetric.}

Considering the types of graphs present at higher loops, each internal strips of the string graphs have their lengths $t_i$ undergoing tropical scaling, $t_i \sim T_i/(2\pi\alpha')$, and the $\J$ cycles contribute contact-terms with width $1/2$. One aspect of the discussion is simpler at higher loops compared to one loop: there are no triangles generated by the internal $\J$ contours with only one puncture. This stems from the fact that the translation invariance which, at one-loop, allowed to fix one point at the origin of the $\mathcal{J}$ cycles is absent at higher loops.\footnote{This also raises the possibility at one loop that a different gauge choice could lead to removing those graphs. But the definition of the loop momentum would be modified and would introduce some extra parameter related to the distance at which the loop momentum is measured relative to an origin of the coordinate system. This parameter would possibly affect the monodromy relations and we do not know what effect it would have in terms of graph representation in the field theory limit.} Therefore, puncture $1$ will not be able to go infinitesimally close to $m$ on a $\J$ cycle, as in figure~\ref{fig:edge_triangle} for instance.\footnote{One could wonder if a configuration where puncture $m$ collides towards the origin of the $\mathcal{J}$ cycles at the same time as puncture $1$ but along a $\mathcal{J}$ cycle, could not lead to a triangle. Such degenerations are of higher order and measure zero at leading order in $\alpha'$.}

\begin{figure}
	\centering
	\includegraphics{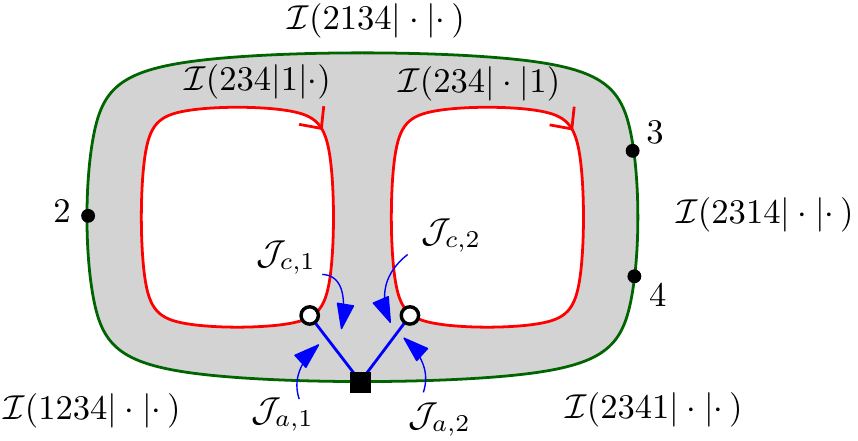}
	\caption{Two-loop  string amplitude with contours from eq.~\eqref{eq:2-loop-monodromy}.}
	\label{fig:two-loop-ct}
\end{figure}

We now consider an explicit example: the planar two-loop monodromy represented by integrating the position of leg $1$ within the two loop surface of figure~\ref{fig:two-loop-ct}. It reads
\begin{multline}
\label{eq:2-loop-monodromy}
\mathcal{I}(1234|\cdot|\cdot)+
e^{i\pi k_1\cdot k_2}\mathcal{I}(2134|\cdot|\cdot)+
e^{i\pi k_1\cdot k_{23}}\mathcal{I}(2314|\cdot|\cdot)+
e^{i\pi k_1\cdot k_{234}}\mathcal{I}(2341|\cdot|\cdot)+\\
e^{i\pi k_1\cdot \ell_1}\mathcal{I}(234|1|\cdot)+
e^{i\pi k_1\cdot \ell_2}\mathcal{I}(234|\cdot|1)=
e^{i\pi k_1\cdot \ell_1}(\mathcal{J}_{a,1}-  \mathcal{J}_{c,1})+  e^{i\pi k_1\cdot \ell_2}(\mathcal{J}_{a,2}-  \mathcal{J}_{c,2}),
\end{multline}
where the various contours are depicted on the figure (for more details see \cite{Casali:2019ihm}). The field theory limit of this relation produces, as usual, two identities, one at order ${\cal O}(1)$ and one at ${\cal O}(\alpha')$. At leading order, the graphs cancel in virtue of the antisymmetry of the $\dot G$ propagator; no contact terms contribute, as they higher order in $\alpha'$. At the next order, the phases produce $\alpha'$ factors and the contact terms do contribute. We focus on graphs involved in the Jacobi identity which flips leg 1 past the origin of the $\J$ cycles on the outer boundary, which is the analogue of the one-loop case we studied above.

\begin{figure}
	\centering
	\includegraphics{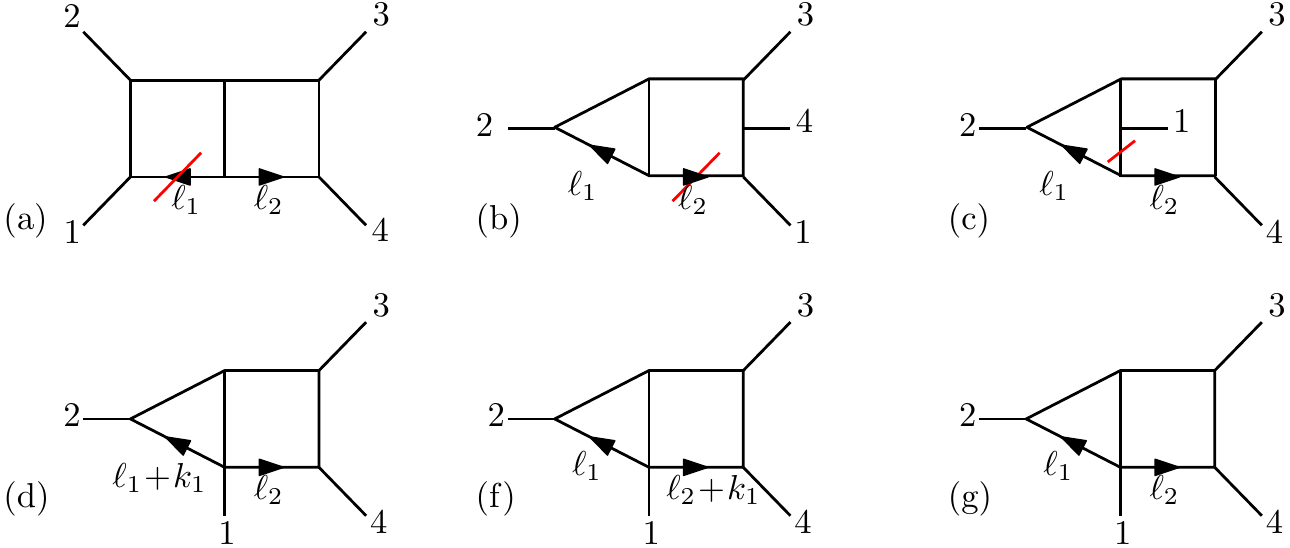}
	\caption{Cancellation systematics at two-loops. Top: two-loop string amplitude with contours from eq.~\eqref{eq:2-loop-monodromy}. Red lines correspond to propagators canceled by the phases of the monodromy relations. Bottom: contact terms generated by the $\J$ contours. While these graphs seem to constitute a Jacobi triplet, this would imply different definitions of the loop momentum than those. Therefore the graphs cannot cancel at fixed loop momentum in the monodromy relations. The role of the contact terms from the $\mathcal{J}_{a/c,1/2}$ integrals is precisely to remove these terms.}
	\label{fig:two-loop-graphs}
\end{figure}

The graphs (a) in figure~\ref{fig:two-loop-graphs} come from the contours $\mathcal{I}(1234|\cdot|\cdot)$ and $\mathcal{I}(234|1|\cdot)$, likewise (b) comes from $\mathcal{I}(2341|\cdot|\cdot)$ and $\mathcal{I}(234|\cdot|1)$ and (c) from $\mathcal{I}(234|\cdot|1)$ and $\mathcal{I}(234|1|\cdot)$. An inspection of the phases in eq.~\eqref{eq:2-loop-monodromy} at order $\alpha'$ shows that correct inverse propagators are reconstructed to cancel the propagators connected at the origin of the $\J$ cycles, as in, e.g., eq.~\eqref{eq:phase_1_j}. This results in three contact-term graphs, which are equal, apart from the important fact that they have different definitions of the loop momenta. Therefore, they cannot cancel by a BCJ mechanism and another contribution from the monodromy relation is needed to remove those graphs.\footnote{Remember that the relations hold at fixed loop-momentum.}

The $\J$ cycles come to the rescue here, once again. Although we have not computed this coefficient from first principle, if the contact they give rise to comes with a coefficient $1/2$, in strict analogy with the one-loop case, the two types of contributions cancel each other out term-by-term, at fixed loop momentum: the contact term $(d)$ comes from $\J_{a,1}$, $(e)$ from $\J_{c,2}$, and $(f)$ from $\J_{c,1}$ and $\J_{a,2}$. Each of these contact terms has the exact same loop-momentum assignment as the graphs with canceled propagators pictured above on figure~\ref{fig:two-loop-graphs}.

Furthermore, one can see that no other region of the moduli space can produce the corresponding terms. The cancellation therefore \textit{must} happen in this way. In this sense we have bootstrapped the field theory limit of the $\J$ contours from the knowledge that the monodromy relations must be satisfied. The only ambiguity that could arise in this bootstrap reasoning relates is the relative coefficient of $\J_{c,1}$ and $\J_{a,2}$ in the sum that cancels the contribution $\mathcal{I}(234|\cdot|1)+\mathcal{I}(234|1|\cdot)$ giving rise to the diagram $(c)$. But this coefficient does not play a role in the field theory limit (only the sum does) so it is guaranteed that $(c)$ and $(g)$ cancel exactly.

\appendix

\acknowledgments
We thank Julio Parra-Martinez for discussions, and the developers of Ipe  (\url{http://ipe.otfried.org}) which was used in creating figures for this paper. We would also like to thank the referee for helpful comments which contributed to a more clear presentation of this work. S.M. gratefully acknowledges the funding provided by Carl P. Feinberg.
This research is supported in part by U.S. Department of Energy grant DE-SC0009999 and by funds provided by the University of California. 

\section{Jacobi theta functions}
\label{sec:conventions}

Here we briefly summarize our conventions used to derive \eqref{eq:G-def} and the following formulae. 
We take the odd Jacobi theta function to be defined by:
\begin{equation}
  \label{eq:theta-1}
    \vartheta_1(z,\tau) = 2q^{1/8}\sin(\pi z)\prod_{n=1}^\infty (1-q^n)(1-q^n w)(1-q^n w^{-1}),
    \end{equation}
    where $w=e^{2 i \pi z}$ and $q=e^{2i\pi\tau}$ is the nome. The first derivative in $z$ evaluated at $z=0$ is then given by
    \begin{equation}
    \vartheta_1'(0,\tau) = 2\pi\eta^3(\tau) = 2\pi q^{1/8} \prod_{n=1}^{\infty}(1-q^n)^3,
\end{equation}
where $\eta(\tau)$ is the Dedekind eta function.
This leads to
\begin{equation}
  \frac{\vartheta_1 (z,\tau)}{\vartheta_1'(0,\tau)} = \frac{\sin(\pi z)}{\pi}\prod_{n=1}^\infty (1-q^n w)(1-q^n w^{-1})(1-q^n)^{-1}.
\end{equation}
Taking a logarithm, performing a Taylor expansion in $q^n$, and collecting constant terms and those that only depend on $\tau$ into a function $c(\tau)$ gives
\begin{equation}
  \ln \frac{\vartheta_1 (z,\tau)}{\vartheta_1'(0,\tau)}  = \log(\sin(\pi z))-\sum_{n=1}^\infty \sum_{m=1}^\infty \frac 1m q^{nm}(w^m+w^{-m}) +c(\tau).
\end{equation}
This is finally summed to
\begin{equation}
  \ln \frac{\vartheta_1 (z,\tau)}{\vartheta_1'(0,\tau)}  = \log(\sin(\pi z))- 2\sum_{m=1}^\infty \frac 1m\frac{ q^{m}}{1-q^m}\cos(2\pi m z) +c(\tau),
\end{equation}
which is the expression used in \eqref{eq:G-def}.

\bibliographystyle{JHEP}
\bibliography{biblio}

\end{document}